\documentclass[aps,pra,epsf,superscriptaddress,amsmath,amssymb,amsfonts,twocolumn,showpacs]{revtex4-2}

\usepackage{graphicx} 
\usepackage[caption=false]{subfig} 
\usepackage{dcolumn}
\usepackage{pict2e}
\usepackage{multirow} 
\usepackage{booktabs}
\usepackage{color} 
\usepackage[dvipsnames]{xcolor} 
\usepackage{amsmath, bm} 
\usepackage{amsfonts, amssymb}  
\usepackage{verbatim}
\usepackage{underscore} 
\newcommand{\ket}[1]{\ensuremath{\vert#1\rangle}}  
\newcommand{\braketoperator}[3]{\ensuremath{\left\langle#1\left\lvert#2\right\rvert#3\right\rangle}} 

\newcommand{\ion}[2]{#1\ensuremath{^{#2+}}}

\begin{document}

\title{State-resolved ionization dynamics of neon atom induced by x-ray free-electron-laser pulses}

\author{Laura Budewig}
\affiliation{Center for Free-Electron Laser Science CFEL, Deutsches Elektronen-Synchrotron DESY, Notkestrasse 85, 22607 Hamburg, Germany}\affiliation{Department of Physics, Universit\"at Hamburg, Notkestrasse 9-11, 22607 Hamburg, Germany} 
\author{Sang-Kil Son}
\affiliation{Center for Free-Electron Laser Science CFEL, Deutsches Elektronen-Synchrotron DESY, Notkestrasse 85, 22607 Hamburg, Germany}
\author{Robin Santra}
\affiliation{Center for Free-Electron Laser Science CFEL, Deutsches Elektronen-Synchrotron DESY, Notkestrasse 85, 22607 Hamburg, Germany}\affiliation{Department of Physics, Universit\"at Hamburg, Notkestrasse 9-11, 22607 Hamburg, Germany}

\date{\today}

\begin{abstract}

We present a theoretical framework to describe state-resolved ionization dynamics of neon atoms driven by ultraintense x-ray free-electron-laser pulses. In general, x-ray multiphoton ionization dynamics of atoms  have been described by time-dependent populations of the electronic configurations visited during the ionization dynamics, neglecting individual state-to-state transition rates and energies. Combining a state-resolved electronic-structure calculation, based on first-order many-body perturbation theory, with a Monte Carlo rate-equation method, enables us to study state-resolved dynamics based on time-dependent state populations.  Our results demonstrate that configuration-based and state-resolved calculations provide similar charge-state distributions, but the differences are visible when resonant excitations are involved, which are also reflected in calculated time-integrated electron and photon spectra. In addition, time-resolved spectra of ions, electrons, and photons are analyzed for different pulse durations to explore how frustrated absorption manifests itself during the ionization dynamics of neon atoms.
\end{abstract}

\maketitle 

\section{Introduction}\label{Introduction}
X-ray free-electron lasers (XFELs)~\cite{Emma,Tanaka,Decking, Kang, Prat} provide x-ray radiation with extremely high intensity and ultrashort pulse durations ranging from a few to a hundred 
femtoseconds~\cite{Pellegrini2}. 
Interaction with these ultraintense XFEL pulses can induce x-ray multiphoton ionization dynamics in matter~\cite{IoIXBwA}. Enabled by the high intensity and, thus, the extremely large number of x-ray photons in a single pulse, multiple sequences of one-photon ionization accompanied by decay processes (Auger-Meitner decay and fluorescence), refilling inner-shell vacancies, can take place. Consequently, atoms or molecules may become highly ionized during interaction with XFEL pulses~\cite{Young2, Fukuzawa,Rudek2, Rudek, Rudenko}.
Such x-ray multiphoton ionization dynamics can be simulated by a rate-equation approach~\cite{Young2, Rohringer,Makris} and were first investigated both experimentally and theoretically in neon atoms~\cite{Young2}. Further studies  for neon have revealed the relevance  of direct nonsequential two-photon ionization in excited neon ions~\cite{Doumy} and resonant excitations at a specific photon energy~\cite{Xiang}.
Moreover, x-ray multiphoton ionization dynamics in heavier atoms~\cite{Fukuzawa, Rudek, Rudek2, Rudek3, Motomura} and molecules~\cite{Rudenko, Berrah, Murphy, Hoener} have been examined in various ways, including resonant effects~\cite{Toyota, Boll, Ho, Ho2}.
Typically, the ionization dynamics of atoms and molecules have been examined with measurement of ions generated after interacting with an intense XFEL pulse, but electron spectra \cite{Mazza,Son2012,Schafer} and photon spectra \cite{Rudek,Son2012,Buth,Schafer} are complementary to ion spectroscopy.
Deepening our understanding of multiphoton ionization dynamics  and the accompanying electronic damage~\cite{Lorenz, Quiney,Son} is relevant for  applications of XFELs, like serial  femtosecond crystallography~\cite{Chapman,Coe} and single particle experiments~\cite{Sobolev, Seibert}, which are limited by electronic damage and structural disintegration of the sample~\cite{Nass}.

Most of the theoretical treatments of x-ray multiphoton ionization dynamics are limited in the way that the states visited during the ionization dynamics are described only by electronic configurations 
 in the rate-equation approach. In this way, transition energies and rates are averaged over individual electronic states for a given electron configuration. 
 As will be discussed in detail in Sec.~\ref{Ionization}, this configuration-based approach already demands to solve a large set of coupled rate equations and the number of rate equations explodes when resonant excitations are included~\cite{Ho,Ho2,Toyota, Rudek}.
 The configuration-based rate-equation approach has been widely used and successfully applied for interpreting and designing many XFEL experiments~\cite{Young2, Fukuzawa, Rudek, Rudek2, Rudenko,  Rohringer, Makris, Doumy,Xiang,Rudek3, Motomura, Berrah, Murphy, Hoener, Toyota, Boll, Ho, Ho2, Son2012, Schafer, Buth,Lorenz, Ciricosta, Buth2, Liu}.
However, it can  neither treat individual state-to-state transitions nor detailed state-resolved ionization dynamics. 

In this work, we investigate x-ray multiphoton ionization dynamics of neon atoms based on individual electronic states by extending the \emph{ab initio} electronic-structure and ionization-dynamics toolkit \textsc{xatom}~\cite{X,Son,XATOM}.
Recently, a state-resolved electronic-structure framework, based on first-order many-body perturbation theory, has been introduced in \textsc{xatom}~\cite{TIofOA}.  
As a follow-up study, we here embed these improved electronic-structure calculations into the  Monte Carlo on-the-fly rate-equation method for describing ionization dynamics~\cite{Son2012, Fukuzawa}. This implementation permits us to perform huge-size rate-equation calculations that are inevitable for state-resolved ionization dynamics calculations. 
We compare both configuration-based and state-resolved approaches for x-ray multiphoton ionization of Ne. 
It can be expected that resonant excitations and spectra should in general profit from our state-resolved implementation for two reasons.
First, as shown in Ref.~\cite{TIofOA}, the first-order-corrected energies, delivered by the improved electronic-structure calculations, often provide better transition energies. Second, individual states associated with a configuration generally do not behave the same during ionization dynamics. Based on the state-resolved approach, we will also investigate the time evolution of charge-state distributions (CSDs) and photoelectron, Auger-Meitner electron, and fluorescence spectra during an XFEL pulse. 

The paper is organized as follows. In Sec.~\ref{theory}, we present our state-resolved Monte Carlo on-the-fly implementation in \textsc{xatom}. Additionally, individual state-to-state resonant photoabsorption cross sections, which are missing in Ref.~\cite{TIofOA}, are addressed.
A comparison with a configuration-based Monte Carlo calculation is the topic of Sec.~\ref{Comparison}, while in Sec.~\ref{Timeevolution}, we study the time evolution of ion, electron, and photon spectra for neon  for different pulse durations. We summarize our findings and discuss future perspectives in Sec.~\ref{conclusion}.

\section{Theoretical details}\label{theory} 

\subsection{Improved electronic-structure calculations}\label{IESC}
Here, we briefly summarize the formalism underlying the improved electronic-structure calculations, implemented in \textsc{xatom}. For more details, the reader is referred to Ref.~\cite{TIofOA}.
\textsc{xatom} is based on the Hartree-Fock-Slater (HFS) approach~\cite{HFS, Son}, in order to keep the calculations feasible and efficient also for heavy atoms and the inclusion of resonant excitations (see Sec.~\ref{Ionization}).
The HFS calculations can be improved through first-order  many-body perturbation theory~\cite{Griffiths:QM,Sakurai:QM} for the full $N$-electronic Hamiltonian~\cite{Griffiths:QM} 
\begin{equation}
\label{eqn:Hmatter}
{\hat{H}}_{\text{matter}} = \sum_{i =1}^{N}\left\lbrace{
-\frac{1}{2}{\nabla}_{i}^2 -\frac{{Z}}{\vert{\vec{x}}_{i}\vert}}\right\rbrace+
\frac{1}{2}\sum_{i \neq j}^{N} \frac{1}{\vert{\vec{x}}_{i}-{\vec{x}}_{j}\vert}.
\end{equation}
Here, ${\vec{x}}_{i}$ is the position of an electron in the atom of nuclear charge $Z$ and atomic units are used.
In this approach, for an electronic configuration of interest, the matrix representation of ${\hat{H}}_{\text{matter}}$ is created in the set of electronic Fock states. The Fock states are antisymmetrized products of spin orbitals that are eigenstates of the HFS Hamiltonian ${\hat{H}}^{\text{HFS}}$~\cite{Gottfried:QM}.
The eigenstates of this matrix provide zeroth-order $LS$ eigenstates $\ket{LSM_LM_S\kappa}$ with first-order-corrected energies $E_{LS\kappa}$, given by the eigenvalues. An important feature of the new states is that they are also eigenstates of the total orbital angular momentum and of total spin. Therefore, they can be labeled by the $L$ and $S$ quantum numbers and their projections $M_L$ and $M_S$, respectively. Since the set of ($L$,$S$,$M_L$,$M_S$) does not always uniquely define the state, we need an additional integer index $\kappa$. Note that the states within a term $^{2S+1}L(\kappa)$ share the same first-order-corrected energy $E_{LS\kappa}$. 
~~\\
\subsection{Individual state-to-state resonant photoabsorption cross sections}\label{CSres}
Having at hand first-order-corrected energies and zeroth-order $LS$ eigenstates, we can perform further state-resolved calculations, e.g., cross sections and transition rates. A detailed description of state-to-state photoionization cross sections, Auger-Meitner decay rates, and fluorescence rates can be found in Ref.~\cite{TIofOA}.

Previous studies of x-ray multiphoton ionization dynamics have demonstrated the importance of resonant excitations at certain photon energies for neon~\cite{Young2, Xiang}, krypton~\cite{Rudek3, Ho2, Ho}, and xenon~\cite{Rudek,Rudek2, Ho, Ho2}. 
Therefore, we present a calculation of individual state-to-state resonant photoabsorption cross sections, based on our improved electronic-structure framework.  It can be performed in a similar way as the calculation of individual state-to-state photoionization cross sections (see Ref.~\cite{TIofOA}).
 
Let us consider a bound-to-bound resonant transition of an electron in the subshell with quantum numbers $(n_h,l_h)$ to a higher-lying $(n_j,l_j)$-subshell by absorbing a linearly polarized photon with a photon energy $\omega$. The associated initial zeroth-order LS eigenstate is \(\ket{{L}_{i}{S}_{i}{M}_{{L}_{i}}{M}_{{S}_{i}}}\) with first-order-corrected energy $E_{L_iS_i}$ and the accessible final target state is  \(\ket{{L}_{f}{S}_{f}{M}_{{L}_{f}}{M}_{{S}_{f}}}\)  with $E_{L_fS_f}$ (here, $\kappa_i$ and $\kappa_f$ are omitted for simplicity).
Then, the state-to-state resonant photoabsorption cross section can be written as  
\begin{widetext}
\begin{equation}
\label{eqn:pcs}
\begin{split}
{\sigma}_{^{2S_i+1}L_i;^{2S_f+1}L_f}^{{M}_{{L}_{i}};{M}_{{L}_{f}}}(n_hl_h,n_jl_j,{\omega})
=&\frac{4\pi^2}{3{\omega}}\alpha\delta(E_{L_fS_f}-E_{L_iS_i}-\omega){({\varepsilon}_{n_hl_h}-{\varepsilon}_{n_jl_j})^2}{l}_{>}\left\vert\int_{0}^{\infty}dr{u}_{{n}_{j} {l}_{j}}^{\ast}(r)r{u}_{{n_h} {l_h}}(r)\right\vert^2\\
&\times\left\vert\sum_{j,h} C({l}_h,{l}_{j},1;{m}_{{l}_{h}},-{m}_{{l}_{j}},0)\braketoperator{{L}_{f}{S}_{f}{M}_{{L}_{f}}{{S}_{f}}}{{\hat{c}}_{j}^\dagger{\hat{c}}_{h}}{{L}_{i}{S}_{i}{M}_{{L}_{i}}{S}_{i}}\right\vert^2.
\end{split}
\end{equation}
\end{widetext}
In this expression, \(C(\cdot)\) represents a Clebsch-Gordan coefficient~\cite{Racah2, Judd:OTiAS} and \({l}_{>} = \text{max}({{l}_{h},{l}_{j}})\). 
The indices $h$ and $j$ denote the involved spin orbitals  in the $(n_hl_h)$- and $(n_jl_j)$-subshells, respectively, between which the electron is transferred.  They have orbital energies $\varepsilon_{n_hl_h}$ and $\varepsilon_{n_jl_j}$, and  quantum numbers $(n_h,l_h,m_{l_h},m_{s_h})$ and $(n_j,l_j=l_h\pm1,m_{l_j}=m_{l_h},m_{s_j}=m_{s_h})$, respectively. The relation between their quantum numbers can be attributed to the selection rules for a dipole transition with linearly polarized photons~\cite{Hertel:AMOP}. Creation and annihilation operators ${{\hat{c}}_{j}^\dagger}$ and ${{\hat{c}}_{h}}$ can be associated with the spin orbitals~\cite{Cixp}.  Since 
the interaction Hamiltonian corresponding to one-photon absorption~\cite{Cixp} does not affect the spin and its projection, the cross section is independent of the initial  and final spin projections.

According to the energy conservation law, the transition energy, $E_{L_fS_f}-E_{L_iS_i}$, equals the photon energy $\omega$,
\begin{equation}
\omega = E_{L_fS_f}-E_{L_iS_i}.
\end{equation}
Therefore, it is critical to take into account that the  pulse has some energy distribution around a given peak photon energy $\omega_\text{in}$. Let us assume a Gaussian pulse profile~\cite{Toyota}
\begin{equation}
G(\omega; \omega_\text{in}, \Delta_{\omega_\text{in}}) = 
\frac{1}{\Delta_{\omega_\text{in}}}\sqrt{\frac{4 \ln 2}{\pi} }e^{- 4 \ln 2   \left( \frac{\omega_\text{in}-\omega}{\Delta_{\omega_\text{in}}}\right)^2 },
\end{equation}
where $\Delta_{\omega_\text{in}}$ is the full width at half maximum (FWHM)  of the bandwidth of the photon energy distribution. 
After employing convolution with the pulse profile, the individual state-to-state  resonant photoabsorption cross section for  a peak photon energy $\omega_\text{in}$  is
\begin{equation}
\label{eqn:csres}
\begin{split}
&{\sigma}_{^{2S_i+1}L_i;^{2S_f+1}L_f}^{{M}_{{L}_{i}};{M}_{{L}_{f}}}(n_hl_h,n_jl_j,\omega_\text{in})
=\\ &{\sigma}_{^{2S_i+1}L_i;^{2S_f+1}L_f}^{{M}_{{L}_{i}};{M}_{{L}_{f}}}(n_hl_h,n_jl_j,\omega=E_{L_fS_f}-E_{L_iS_i})\\ &\times G(E_{L_fS_f}-E_{L_iS_i}; \omega_\text{in}, \Delta_{\omega_\text{in}}).
\end{split}
\end{equation}
Here, ${\sigma}_{^{2S_i+1}L_i;^{2S_f+1}L_f}^{{M}_{{L}_{i}};{M}_{{L}_{f}}}(n_hl_h,n_jl_j,\omega=E_{L_fS_f}-E_{L_iS_i})$ is given by Eq.~(\ref{eqn:pcs}), but without the delta distribution that vanishes due to integration over $\omega$.

It is also worthwhile to mention that 
calculating resonant photoexcitation cross sections requires additional computational parameters.
First, in order to keep the calculation feasible, the number of allowed $(n,l)$-subshells that can be resonantly excited has to be restricted  and to be checked for computational convergence.
In Sec.~\ref{Comparison}, we have used a maximal $n$ quantum number of $n_{\text{max}}=7$ and a maximal $l$ quantum number of $l_{\text{max}}=2$ for Ne calculation at a photon energy of 1050~eV. 
Second, in order to guarantee an accurate description of higher-$n$ states the maximum radius employed in the numerical calculation of orbital and orbital energies~\cite{Son,X} has to be sufficiently large for both bound and continuum states. We used $r_\text{max}$=100~a.u.\ for both in the following resonance-related calculations. Third, due to the convolution procedure, it is necessary to scan the transition energy to search for all accessible resonant bound-to-bound transitions. However, the state-resolved calculation of first-order-corrected transition energies  costs a considerable amount of computational time, when a lot of open subshells are included. For instance, calculating all cross sections for initial \ion{Ne}{3} ($1s^02s^22p^44p^1$) costs roughly two minutes in CPU time, due to the scanning of all resonant excitations with $n_\text{max}\leq7$ and $l_\text{max}\leq2$ (for comparison the nonresonant calculation of photoionization cross sections takes less than three seconds).  
Therefore, when we search for accessible resonant states, we employ an energy scan criterion based on the zeroth-order transition energies, instead of the first-order-corrected values. In our calculation for Ne at 1050~eV, a bandwidth of 1\% of the photon energy was considered. Then, we scan $\pm$157.5~eV ($\pm$15\% of the photon energy) to pick up resonant transitions based on the zeroth-order transition energy. Only for these resonant transitions we consider whether the first-order transition energy lies within the photon energy bandwidth. This reduces the computational time for the upper example to thirty seconds. Note that the differences between zeroth- and first-order transition energies are less than 50~eV in our calculations (see Table~\ref{tab:Res}).

\subsection{State-resolved ionization dynamics calculations }\label{Ionization}
X-ray multiphoton ionization dynamics can be described by a rate equation approach~\cite{Young2, Rohringer,Makris}. In this approach, the time evolution of the population  $P_I(t)$ of a state $I$ is given by a set of coupled rate equations,
\begin{equation}
\label{eqn:RQ}
\frac{dP_I(t)}{dt} = \sum_{I^{\prime}\neq I}\left[{\Gamma}_{I^{\prime}\rightarrow I}P_{I^{\prime}}(t)-{\Gamma}_{I \rightarrow I^{\prime}}P_I(t)\right],
\end{equation}
for all possible states $\lbrace I \rbrace$.
In this expression, ${\Gamma}_{I \rightarrow I^{\prime}}$ is the rate for a transition from the state $I$ to the state $I^{\prime}$ via either photoionization, photoexcitation, or  relaxation (i.e., Auger-Meitner decay or fluorescence). In a configuration-based approach  $\{ I \}$  are defined by all possible electronic configurations, whereas in our state-resolved approach $\{ I \}$ are defined by the electronic configurations together with  additional quantum numbers needed for specifying zeroth-order $LS$ eigenstates ($L$, $S$, $M_L$, $M_S$, $\kappa$).

There are two ways for solving the set of coupled rate equations of Eq.~(\ref{eqn:RQ}): either directly with precalculated rates and cross sections~\cite{Son} or via a Monte Carlo method~\cite{Son2012}. The latter has been extended to a more efficient on-the-fly approach~\cite{Fukuzawa}, i.e., quantities are only calculated when needed, which will be further explained in the following subsection.  The Monte Carlo on-the-fly method is especially crucial for heavy atoms, for which the number of coupled rate equations to be solved becomes extremely large. The number of coupled rate equations is equivalent  to the number of all possible  states or all possible configurations, depending on the definition of $\{I\}$.   

For the configuration-based approach, an estimate of this number can be found in Ref.~\cite{Toyota} (also see Ref.~\cite{Ho} for the case of resonant excitations). Let us consider an initial electronic configuration, $1s^{N_1} 2s^{N_2} 2p^{N_3} 3s^{N_4} 3p^{N_5} \cdots$, where the number of electrons is given by $N_\text{elec} = \sum_j^\text{all} N_j$. During x-ray multiphoton ionization dynamics, different electronic configurations can be constructed by adding zero, one, or up to $N_j$ electrons for each $j$th subshell.  Here, the index $j$ labels all subshells that can be ionized by the given photon energy via one-photon ionization and
no resonant excitation is assumed. The number of all possible configurations is then evaluated by
\begin{equation}
N_\text{config} = \prod_{j} \left( N_j + 1 \right).
\end{equation}
For example, Ne has $1s^2 2s^2 2p^6$ and $N_\text{config} = 3 \times 3 \times 7 = 63$ if all subshells are accessible for one-photon ionization.

For the state-resolved approach, the number of possible zeroth-order $LS$ eigenstates (equal to the number of electronic Fock states) can be estimated as follows.
For each $j$th subshell with $(n,l)$, there are $N_j^\text{max} (= 4 l + 2)$ spin orbitals with different $m_l (\in \lbrace-l, -l+1,\cdots ,l-1,l\rbrace)$ and $m_s (= \pm \frac{1}{2})$, which equals the maximum occupancy. By adding zero, one, or up to $N_j$ electrons in the $j$th subshell, the number of possible states is given by the sum of the numbers of possibilities to distribute added electrons into $N_j^\text{max}$ spin orbitals (each spin orbital has a maximal occupation number of one),
\begin{equation}
\label{eqn:X}
N_\text{state}^j = \sum_{k=0}^{N_j} { N_j^\text{max} \choose k },
\end{equation}
where ${ a \choose b}$ is a binomial coefficient. Then, the number of all possible states is given by multiplying the $N_\text{state}^j$ for all $j$ (no resonant excitation is considered),
\begin{equation}
N_\text{state} = \prod_j \sum_{k=0}^{N_j} { N_j^\text{max} \choose k }.
\end{equation}
If we consider the ground-state configuration, all subshells are fully occupied ($N_j = N_j^\text{max}$), except for the outermost shell (index $j^\prime$  in what follows), which may be partially occupied. For a closed subshell ($N_j = N_j^\text{max}$), $N_\text{state}^j = 2^{N_j}$. Thus, the number of all possible states is written as
\begin{equation}
\label{eqn:Nstates}
N_\text{state} = \left[ \prod_{j \neq j'} 2^{N_j} \right] \cdot \left[ \sum_{k=0}^{N_{j'}} { N_{j'}^\text{max} \choose k } \right].
\end{equation}
If the system has no partially occupied subshells initially and all the subshells are accessible for one-photon ionization, then it is further simplified to $N_\text{state} = 2^{N_\text{elec}}$. For example, Ne has 10 electrons and $N_\text{state} = 2^{10} = 1024$.

If resonant excitations are taken into account, a similar expression to Eq.~(\ref{eqn:X}) can be directly used. Let $N_\text{so}$ be the number of available spin orbitals given by computational parameters $n_\text{max}$ and $l_\text{max}$, and $N_\text{elec}$ be the number of accessible electrons for one-photon ionization or resonant excitation. Then, the number of states is given by
\begin{equation}
N_\text{state}^\text{res} = \sum_{k=0}^{N_\text{elec}} { N_\text{so} \choose k }.
\end{equation}

\begin{figure}[tbp]  
\centering
\includegraphics[trim = 5mm 142mm 5mm 0mm,  clip, width=1\linewidth]{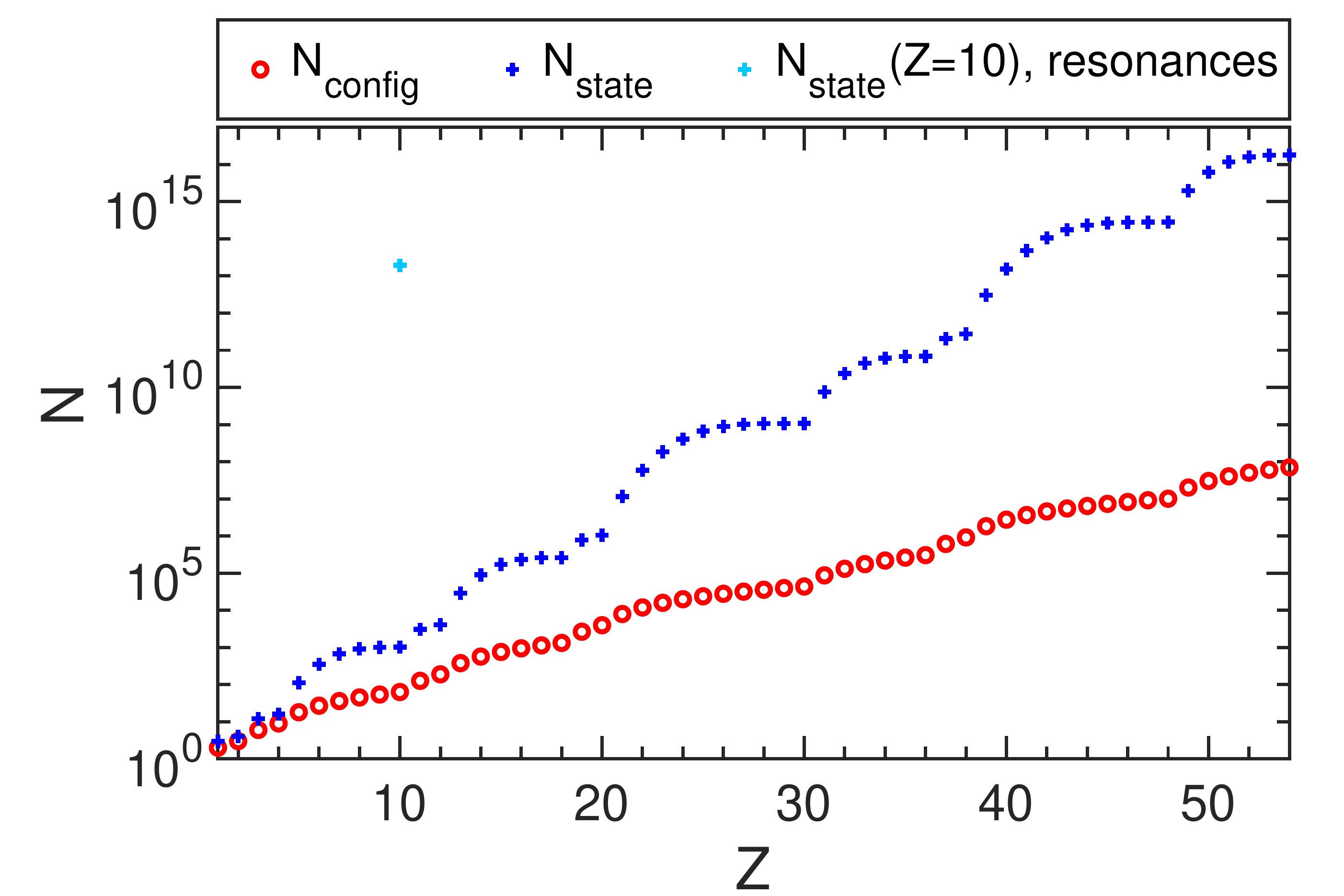}
\caption{Number of configurations $N_{\text{config}}$ (red dots) and number of states $N_{\text{state}}$ (blue crosses) as a function of the nuclear charge $Z$, assuming that all electrons are actively involved in the ionization dynamics. Only the nonresonant case is shown and the ground-state configurations are given by the Aufbau principle. For the resonant case, solely the point of $N_{\text{state}}$ for neon ($Z=10$) with  $n_{\text{max}}=7$ and $l_{\text{max}}=2$ is marked (lightblue).
 }
\label{fig:N}
\end{figure}
Figure~\ref{fig:N} shows the number of configurations and states as a function of the nuclear charge $Z$ for the nonresonant cases. The ground-state electronic configurations are constructed by the Aufbau principle. For all $Z$ the photon energy is assumed to be large enough to ionize all subshells including the $1s$ subshell.
Both $N_\text{config}$ and $N_\text{state}$ grow exponentially, but $N_\text{state}$ is much larger than $N_\text{config}$ for a given $Z$. For the resonant case, this number explodes even for a low-$Z$ system like Ne with limited computational parameters ($n_\text{max}$=7 and $l_\text{max}$=2; $N_\text{so}$=100). With them, $N_\text{state} \simeq 1.9\times10^{13}$, which is marked in Fig.~\ref{fig:N}. Therefore, even for low $Z$ it is inevitable to employ a Monte Carlo on-the-fly scheme for state-resolved ionization dynamics when including resonant excitations.

\subsection{State-resolved Monte Carlo implementation}\label{StateresolvedMC}
In the state-resolved approach, the number of coupled rate equations [Eq.~(\ref{eqn:RQ})] that have to be solved is equal to the number of states [Eq.~(\ref{eqn:Nstates})], as depicted in Fig.~\ref{fig:N}.
We implement a state-resolved Monte Carlo on-the-fly algorithm within \textsc{xatom}~\cite{XATOM}. This allows us to apply our state-resolved ionization dynamics framework to heavier atoms, like argon ($Z=18$) or xenon ($Z=54$), and to the resonant case.

In general, in a Monte Carlo approach for ionization dynamics, we stochastically consider many trajectories for possible ionization pathways, i.e., sequences of repeated one-photon ionization and inner-shell relaxation events. The populations of entities, such as charge state, electronic configuration, or electronic state, are then obtained by averaging over an ensemble of trajectories. A detailed description of the configuration-based Monte Carlo method can be found in Ref.~\cite{Son2012}. Extending it to a state-resolved  Monte Carlo algorithm  basically requires to replace a configuration index with a combination of configuration and state indexes through the whole algorithm, i.e., $I = I_{\text{config}}\rightarrow I=\left(I_{\text{config}},I_{LS}\right)$. Here, $I_{\text{config}}$ indicates an electronic configuration and $I_{LS}$ indicates the additional quantum numbers needed for specifying a zeroth-order $LS$ eigenstate. 
Note that we do not include in $I_{LS}$  the spin projection $M_S$ and, hence, do not distinguish between states with different spin projection. Because of a lack of spin coupling for all involved interaction Hamiltonians, states with different spin projections always have the same transition probabilities and, consequently, behave exactly the same during the ionization dynamics.  Thus, $M_S$ can be neglected in the description of the individual states.
Moreover, cross sections  and rates based on configurations need to be replaced by  individual state-to-state cross sections and  rates~\cite{TIofOA}.

For the sake of completeness, we sketch our state-resolved Monte Carlo on-the-fly implementation:

(a) Start with the initial electronic configuration $I_{\text{config}}$, i.e., that for the neutral atom, and calculate all zeroth-order $LS$ eigenstates  for the initial configuration via the improved electronic-structure implementation (see Sec.~\ref{IESC} and Ref.~\cite{TIofOA}). If there is more than one $LS$ eigenstate, the state with the minimal first-order-corrected energy $E_{LS\kappa}$ is selected. If  $L \neq 0$, the $M_L$ projection quantum number is randomly chosen as an initial condition for each trajectory. If $S\neq 0$, then the maximal spin projection is chosen for convenience (it does not influence the ionization dynamics). In this way we set up the initial state $I=\left(I_{\text{config}},I_{LS}\right)$. In order to reduce the computational effort, store the information about the electronic structure, so that it can be directly reused for further trajectories. 

(b) Set up an initial value for the time $t$ and the time step $\Delta_t$.

(c) Calculate all individual state-to-state cross sections $\sigma_k$ and transition rates $\Gamma_k$ for all transitions from the current state $I=\left(I_{\text{config}},I_{LS}\right)$ to the accessible  final state $I^k=\left(I_{\text{config}}^k,I_{LS}^k\right)$. Transition energies are also calculated based on the first-order-corrected energies for the current state, i.e., $E_{LS\kappa}$, and the final state, i.e., $E_{L_kS_k\kappa_k}$. Note that we employ the same orbital set optimized for the current (initial) state to evaluate the final state energy~\cite{TIofOA}. 
Here, $k$ is used as a label for  one of the processes, running  from 1 to the number of all possible individual state-resolved transitions. The index $k$ in $L_k$, $S_k$,  etc.\ indicates that the new state is reached via the $k$th process. Also the information about the possible processes is stored for  later use. 

(d) Cross sections and rates  determine the transition probability $p_k$ at time $t$ for the $k$th process via 
\begin{equation}
\label{eqn:pk}
p_k =
\begin{cases} \Gamma_k \Delta_t & \text{for decay process},\\
 \sigma_k J(t) \Delta_t & \text{for photoionization/absorption},
 \end{cases}
\end{equation}
with $J(t)$ being the time-dependent photon flux.

(e) Select a process $k$ randomly,  taking into consideration the different transition probabilities $p_k$ of all possible processes (for more details, see Ref.~\cite{Son2012}).

(f) Counters for the time-dependent charge-state distribution (CSD) and time-resolved spectra are considered as follows. The time- and energy-bin counter for the photoelectron, the Auger-Meitner electron, or the fluorescence photon is increased by one, according to the electron kinetic energy or emitted photon energy of the selected $k$th process  at a given time $t$. This counter will be used for generating time-resolved spectra in Sec.~\ref{Timeevolutionspectra}. Integrating this counter over time corresponds to the spectra after time evolution shown in Sec.~\ref{ComparisonSpectra}. Regarding time-dependent CSDs in Sec.~\ref{TimeevolutionCSD}, the charge state of the given $I$ is examined at every time step of time bins and the corresponding counter is increased by one. 

(g) Continue by proceeding to the $k$th process. Set up the new electronic configuration, i.e, $I_\text{config} = I_\text{config}^k$, and the new zeroth-order $LS$ eigenstate $\ket{L_kS_kM_{L_k}\kappa_k}$, i.e., $I_{LS}= I_{LS}^k$. To get proper first-order-corrected energies, a new electronic-structure calculation has to be performed and stored.

(h) Set up a new $\Delta_t$ based on the total transition probability, i.e., the sum over all $p_k$'s. The new $\Delta_t$ is smaller than or equal to the initial $\Delta_t$ chosen in (b). The updated $\Delta_t$ should guarantee that the total transition probability stays significantly smaller than unity when the updated $\Delta_t$ is used.  Then go to the next time step $t\rightarrow t + \Delta_t$.

(i) Repeat the time evolution [(c)--(h)], until the total transition probability is zero. When no process is available any longer, the Monte Carlo trajectory ends. The counter for the final charge is increased by one, which will produce asymptotic CSDs in Sec.~\ref{ComparisonCSD}.

(j) Run many more trajectories [(a)--(i)] and at every 100 trajectories, or in principle any other step size, check whether the final CSD is converged. If this is the case, stop the Monte Carlo iteration.

(k) Results for  ensemble-averaged  CSDs and spectra are obtained by dividing the corresponding counters by the total number of trajectories.

In what follows, we employ a maximal number of Monte Carlo trajectories of 100000 and  a minimal CSD convergence of $10^{-4}$ for asymptotic results in Sec.~\ref{Comparison} and $5 \times10^{-5}$ for time-resolved results in Sec.~\ref{Timeevolution}, respectively. The actual numbers of Monte Carlo trajectories, being necessary to achieve the demanded convergence, range from 12500 to 50300.  For time-independent spectra, 1-eV bins are used, whereas 2-eV bins are used for the time-resolved spectra. For the time-resolved CSD and  spectra, fixed temporal bins are chosen according to the pulse duration.

\section{Asymptotic results }\label{Comparison}
We first contrast  the configuration-based and state-resolved Monte Carlo methods regarding temporally asymptotic results for neon, i.e.,  CSDs (Sec.~\ref{ComparisonCSD}) and electron and photon spectra (Sec.~\ref{ComparisonSpectra}) at the end of time evolution when the pulse is over and all decay processes are completed. 
The state-resolved calculation is performed with the present implementation, whereas for the configuration-based  calculation, we employ the original version of \textsc{xatom}.
We do not include direct nonsequential two-photon absorption~\cite{Doumy, Sytcheva} because its contribution is much smaller than one-photon absorption if the latter is available (for example, inner-shell nonsequential two-photon absorption vs.\ valence one-photon absorption). Above-threshold ionization in the x-ray regime~\cite{Tilley} is also negligible  in the range of intensities that current XFEL facilities can produce.
For simplicity, we also do not include higher-order many-body processes such as double photoionization~\cite{Schneider} and double Auger-Meitner decay~\cite{Kolorenc} via shakeoff and knockout mechanisms. Note that the branching ratio of double photoionization after Ne $K$-shell photoabsorption is about $23\%$ and that of double Auger-Meitner decay is about $6\%$~\cite{Saito}. Including shakeoff processes in the rate-equation approach markedly improves comparison with experimental CSDs~\cite{Doumy, Buth}, especially regarding the odd-even charge-state relation. Finally, the rate-equation approach employed here does not capture coherent effects such as Rabi flopping~\cite{Rohringer2, Rohringer3, Kanter, Li, Nandi,Cavaletto}. For a stochastic ensemble of XFEL pulses based on the self-amplified spontaneous emission principle, these are minor effects \cite{Rohringer2,Rohringer3,Kanter}.

For x-ray beam parameters, we use a  temporal Gaussian pulse envelope  with  10~fs FWHM and a fluence  of $10^{12}~\text{photons}/\mu \text{m}^2$. Note that these are typical x-ray parameters at current XFEL facilities~\cite{Emma, Tanaka,Decking}. The volume integration~\cite{Toyota2}, which is necessary for  quantitative comparison with experimental data, is not considered here. Following Ref.~\cite{Young2},
 three different  photon energies are examined: (i) 800~eV is below the $1s$ threshold of neutral Ne and all Ne ions, (ii) 1050~eV lies in the middle of the $1s$ threshold region of a series of Ne ions, and (iii) 2000~eV is above the $1s$ threshold of all Ne ions. Note  that in the case~(ii) resonant excitations for some transiently formed ions play a relevant role~\cite{Xiang}. Thus, we include resonant bound-to-bound  excitations in this case. 

The computational time was about 6 minutes for 800~eV (21000 Monte Carlo trajectories), about 13 minutes for 2000~eV (50000 trajectories), and about 7 hours for 1050~eV (18400 trajectories) on an Intel Xenon E5-2609 CPU (single core).

\subsection{Comparison of charge-state distributions }\label{ComparisonCSD}
\begin{figure}[tbp]  
\centering
\includegraphics[trim = 10mm 5mm 10mm 0mm,  clip, width=\linewidth]{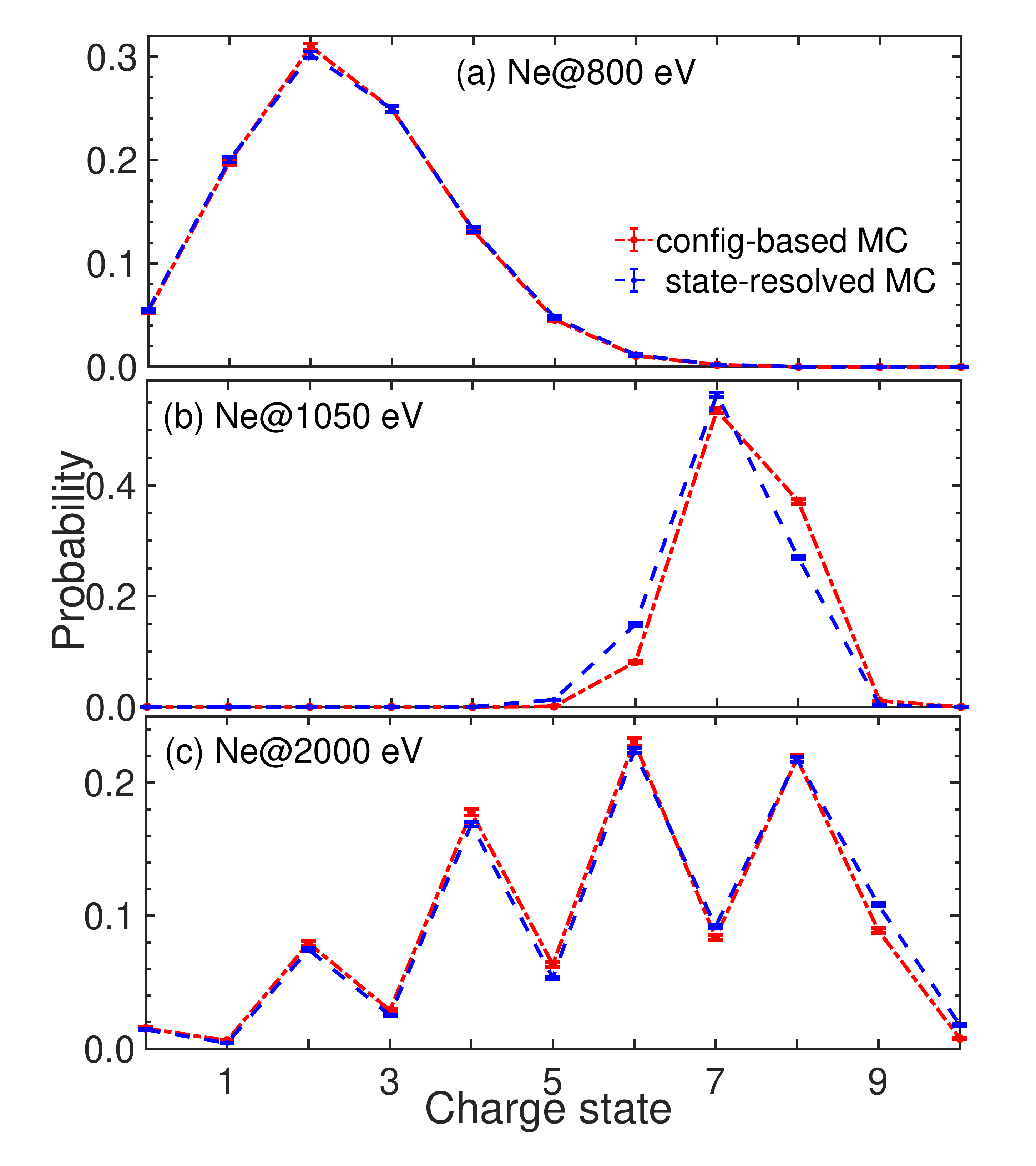}
\caption{Comparison of Ne CSDs obtained with the configuration-based (red) and the state-resolved (blue) Monte Carlo calculations: (a) 800~eV, (b) 1050~eV, and (c) 2000~eV. In all cases,  the Gaussian-shaped pulse has a duration of 10 fs FWHM and a fixed  fluence of $F=10^{12}~\text{photons}/ \mu \text{m}^2$ is used. For (b), resonant excitations up to $n_{\text{max}}=7$ and $l_{\text{max}}=2$ are considered and an energy bandwidth of $1\%$ is assumed. The error bar indicates the statistical error.
}
\label{fig:Ecomparison}
\end{figure}

Figure~\ref{fig:Ecomparison} presents Ne CSDs at the three different photon energies. The population probability $P_q$ of the charge state $q$ is given by the sum of all $P_I$'s (configurational population or state population) belonging to $q$. 
The error bars represent the statistical error estimate ~\cite{Efron:BT}  for each charge state $q$, given by
$\epsilon_{q}= \sqrt{P_q(1-P_q) / (N_\text{traj}-1)} $,
 where $N_\text{traj}$ is the number of Monte Carlo trajectories.
 Comparison in Fig.~\ref{fig:Ecomparison} shows that the state-resolved  calculation is in overall good agreement with the configuration-based calculation, in particular, when the photon energy is off resonance [Figs.~\ref{fig:Ecomparison}(a) and \ref{fig:Ecomparison}(c)]. At 2000~eV, population probabilities differ beyond the error bars only for high charge states. This can be explained by slightly higher transition probabilities in the state-resolved approach caused by the use of first-order-corrected energies and the appearance of a generally nonuniform distribution of individual states for the intermediate configurations.
 
On the other hand, the differences between the two approaches are noticeable when resonant excitations play a role [Fig.~\ref{fig:Ecomparison}(b)]. 
Resonant photoexcitation cross sections are sensitive to the differences between calculated transition energies and the given photon energy. As a consequence, different resonant excitations can be encountered in the state-resolved and configuration-based ionization dynamics calculations (see Table~\ref{tab:Res} in Appendix~\ref{Tables}). For example, the production of \ion{Ne}{8} is enhanced in the configuration-based calculation at the expense of suppression of \ion{Ne}{6}. More detailed analyses regarding relevant resonances  are provided later when electron and photon  spectra are discussed  in Sec.~\ref{ComparisonSpectra}. 

\subsection{Comparison of electron and photon spectra }\label{ComparisonSpectra}

\begin{figure}[tbp]  
\centering
\includegraphics[trim = 10mm 100mm 5mm 0mm,  clip,  width=\linewidth]{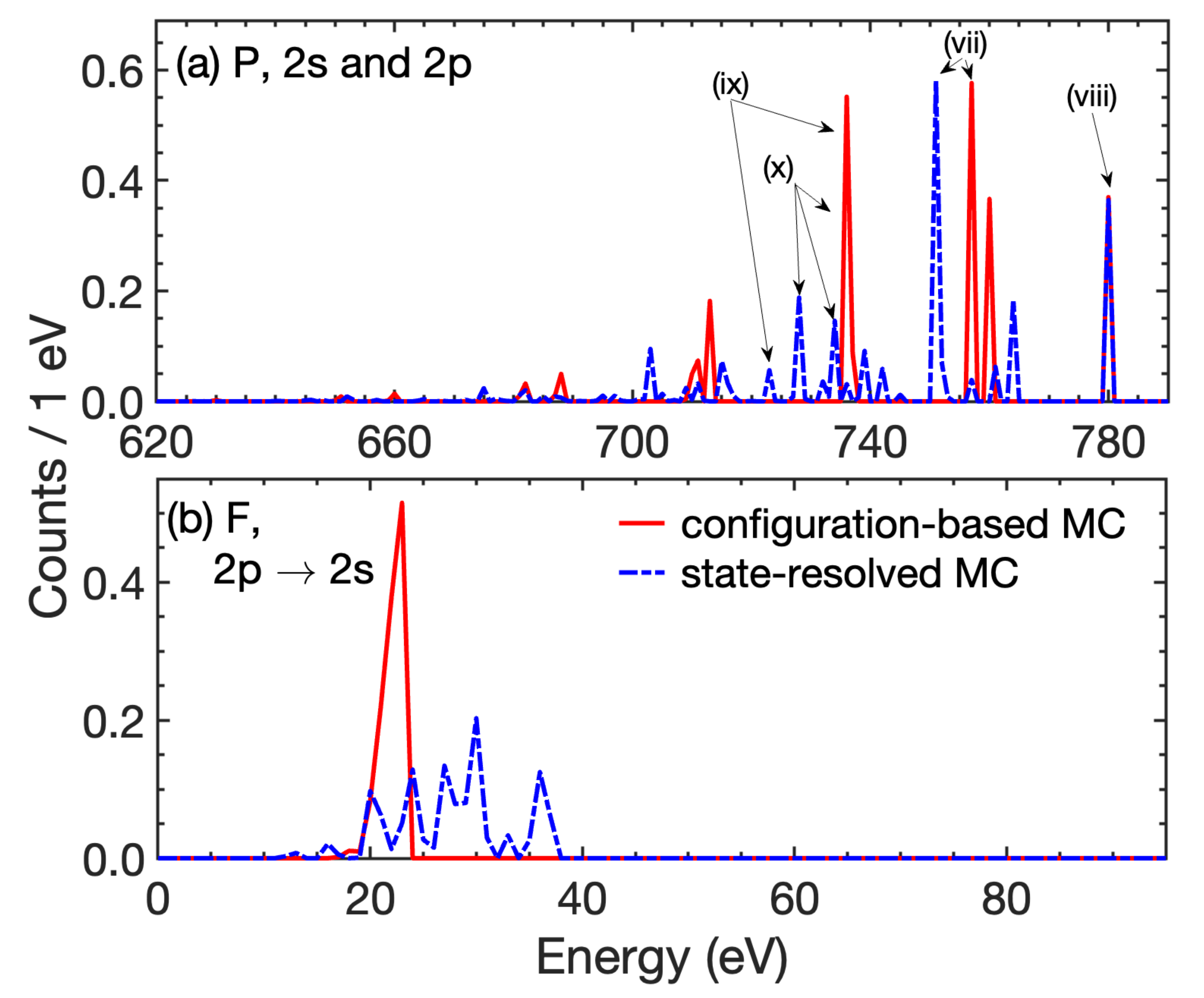}
\caption{(a) Photoelectron (P) and (b) fluorescence (F) spectra of Ne at a photon energy of 800~eV. Other x-ray parameters are the same as used in Fig.~\ref{fig:Ecomparison}. The peak labels in (a) are explained in Table~\ref{tab:P} in Appendix~\ref{Tables}. }
\label{fig:Spectra1}
\end{figure}
\begin{figure}[htbp]  
\centering
\includegraphics[trim = 10mm 0mm 5mm 0mm,  clip,  width=\linewidth]{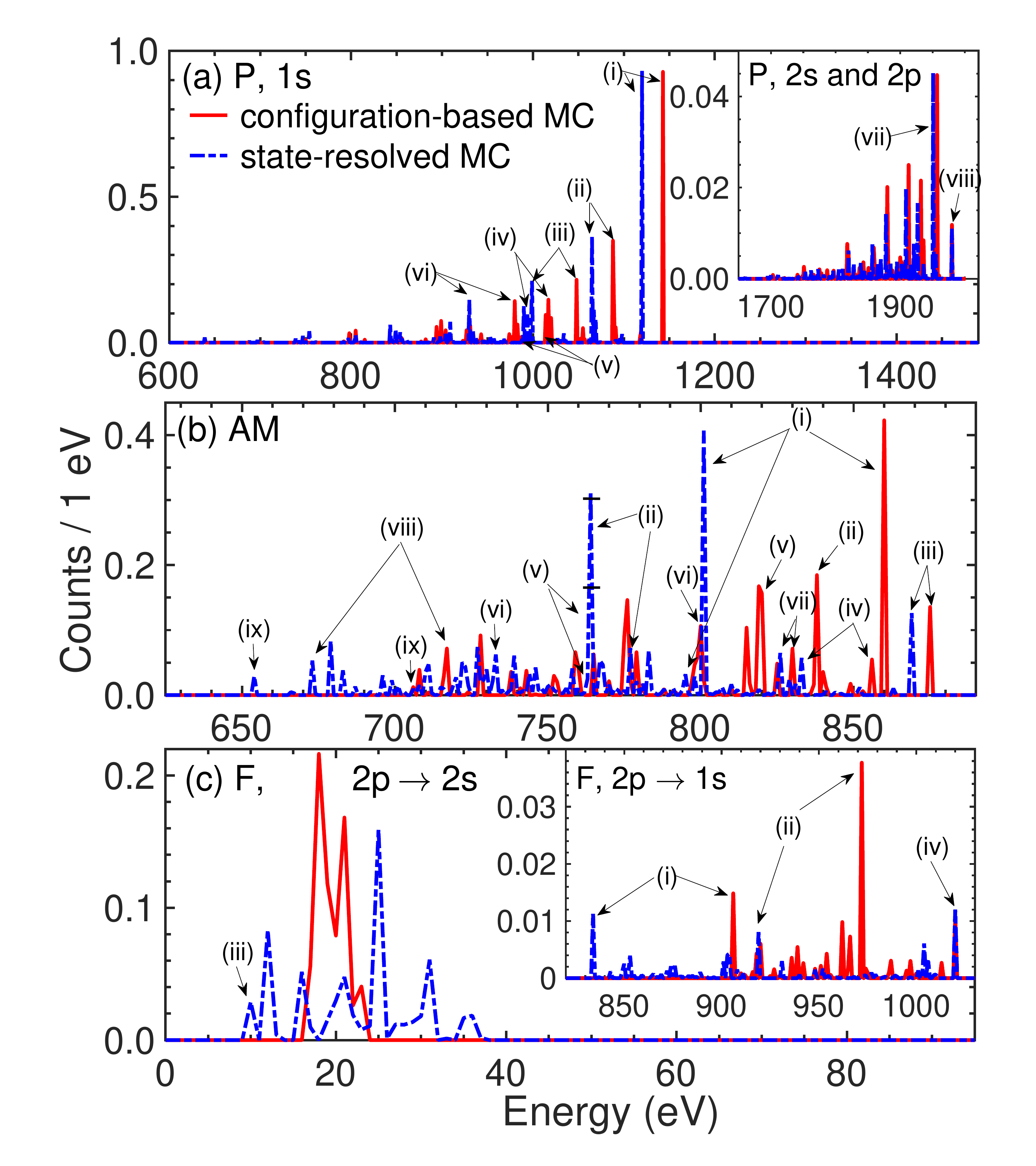}
\caption{(a) Photoelectron (P), (b) Auger-Meitner electron (AM), and (c) fluorescence (F) spectra of Ne at a photon energy of 2000~eV. Other x-ray parameters are the same as used in Fig.~\ref{fig:Ecomparison}. The peak labels are explained in Tables~\ref{tab:P}--\ref{tab:F} in Appendix~\ref{Tables}.}
\label{fig:Spectra3}
\end{figure}
\begin{figure}[htbp]  
\centering
\includegraphics[trim = 10mm 0mm 5mm 0mm,  clip,  width=\linewidth]{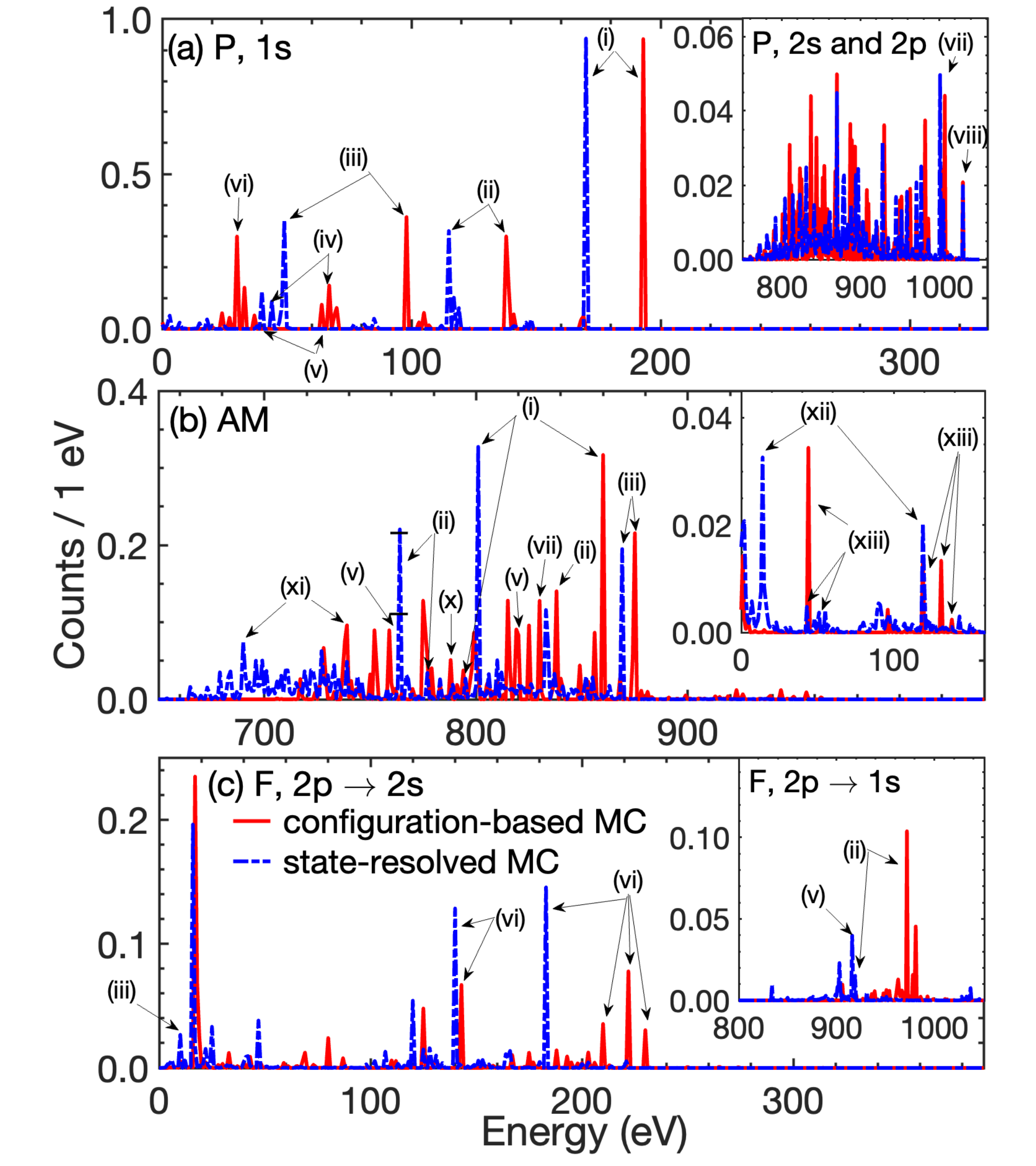}
\caption{(a) Photoelectron (P), (b) Auger-Meitner electron (AM), and (c) fluorescence (F) spectra of Ne at a photon energy of 1050~eV. Other  parameters are the same as used in Fig.~\ref{fig:Ecomparison}. The peak labels are explained in Tables~\ref{tab:P}--\ref{tab:F} in Appendix~\ref{Tables}.   }
\label{fig:Spectra2}
\end{figure}

Figure~\ref{fig:Spectra1}  shows (a) photoelectron and (b) fluorescence spectra at an incoming photon energy of 800~eV and compares the state-resolved (blue dashed line) and configuration-based (red solid line) calculations. At this photon energy, $1s$ ionization is not available, so there is no Auger-Meitner spectrum. In the photoelectron spectrum in Fig.~\ref{fig:Spectra1}(a), some of the dominant peaks are labelled with roman numbers; the corresponding physical processes are specified in Table~\ref{tab:P} in Appendix~\ref{Tables}.
The configuration-based approach employs transition energies computed from zeroth-order energies, i.e., the sum of orbital energies according to the involved configurations. On the other hand, in the state-resolved approach, transition energies are computed based on the first-order-corrected energies for the initial and final states. The energy levels that are degenerate in zeroth-order energies split up in first-order many-body perturbation theory. Consequently, peaks in the state-resolved spectra are not only shifted, but spectra are also broadened with more peaks.  
The energy shifts are clearly visible in Fig.~\ref{fig:Spectra1}(a), except for (viii) $2p$ ionization where two peaks coincide. Splittings also clearly manifest in the photoelectron lines [e.g., peaks (x) in Fig.~\ref{fig:Spectra1}(a)] and in the fluorescence spectra of  Fig.~\ref{fig:Spectra1}(b), where in the configuration-based spectra many peaks coincide around 22~eV.
Note that this behavior of energy shifts and splittings in the state-resolved spectra is a general feature, so it can be found at other photon energies as will be shown below.
 
Figure~\ref{fig:Spectra3} shows (a) photoelectron, (b) Auger-Meitner electron, and (c) fluorescence spectra at 2000~eV with the same x-ray beam parameters as used for Fig.~\ref{fig:Ecomparison}(c). The energy shifts and splittings with the state-resolved approach are clearly exhibited in the Auger-Meitner spectrum. Since the Auger-Meitner peaks in Fig.~\ref{fig:Spectra3}(b) are not well separated and they merge within a narrow energy window, resulting in a complex spectrum, it is critical to apply improved transition energy calculations to interpret individual peaks.
For the single-core-hole Auger-Meitner line (i), the state-resolved result shows considerable improvement towards experimental data as demonstrated in Ref.~\cite{TIofOA}. The energy shift from the configuration-based result to the state-resolved result is $-59$~eV. For the double-core-hole Auger-Meitner line (iii), which is also called $KK-KLL$ hypersatellite~\cite{Southworth}, the energy shift is somewhat smaller ($-6$~eV). Even in this case, the state-resolved value (868.84~eV) is closer to the experimental values (870.50~eV~\cite{Southworth} and 870~eV~\cite{Goldsztejn}), than the configuration-based value (875.27~eV). Note that the prominent peak of the state-resolved approach at 764~eV in Fig.~\ref{fig:Spectra3}(b) is the sum of Auger-Meitner lines (ii) and (v), and other minor contributions that are not assigned here. 
Regarding the fluorescence spectra,  peak  (ii) in Fig.~\ref{fig:Spectra3}(c) is considerably reduced in the state-resolved approach. This is because the initial configuration of (ii) has two states, $^1P$ and $^3P$ (see Table~\ref{tab:F} in Appendix~\ref{Tables}), and the latter cannot relax via $2p \rightarrow 1s$ fluorescence (final state: $^1S$) since a triplet to singlet transition is forbidden in a nonrelativistic calculation. Once the triplet initial state (Ne$^{8+}$ $1s^1 2p^1\ ^3P$) is formed during the state-resolved dynamics, it has to relax via $2p \rightarrow 2s$ fluorescence, giving rise to peak (iii) in Fig.~\ref{fig:Spectra3}(c). Thus, the changes of the peak heights provide more details about underlying physical processes between state-resolved and configurations-based ionization dynamics. The fluorescence peak positions of (iv) in Fig.~\ref{fig:Spectra3}(c) coincide for both approaches.

In Fig.~\ref{fig:Spectra2}, we investigate the effects of resonant excitations on  the electron and photon spectra at 1050~eV. 
The x-ray beam and computational parameters are the same as used in Fig.~\ref{fig:Ecomparison}(b). In the state-resolved and configuration-based approaches, different resonant excitations are predominantly involved in the ionization dynamics at 1050 eV owing to different transition energy calculations (see Table~\ref{tab:Res} in Appendix~\ref{Tables}). The different resonant excitations are all reflected in the spectra in Fig.~\ref{fig:Spectra2}.  For example, photoelectron peak (vi) in Fig.~\ref{fig:Spectra2}(a), which is prominent in the configuration-based approach, is absent  in the state-resolved approach. This is because (vi) refers to the $1s$ ionization of Ne$^{3+}$ $1s^1 2l^6$ $(l=s,p)$ and its threshold is higher than the photon energy in the state-resolved approach.
Instead, resonant excitation of single-core-excited \ion{Ne}{3} predominantly via a $1s\rightarrow4p$ resonant transition is the alternative process (see Table~\ref{tab:Res} in Appendix~\ref{Tables}).
The same transition can also occur at Ne$^{6+}$ $1s^2 2l^2$ within the state-resolved approach. These $1s \rightarrow 4p$ transitions at Ne$^{3+}$ and Ne$^{6+}$ are responsible for the Auger-Meitner decay involving $4p$, which explains the emergence of (xii) in Fig.~\ref{fig:Spectra2}(b), only in the state-resolved approach. On the other hand, in the configuration-based approach, the $1s \rightarrow 3p$ transition is dominant at Ne$^{5+}$ $1s^1 2l^4$ and the resulting double-core-hole-excited state of Ne$^{5+}$ $1s^0 2l^4 3p^1$ relaxes via  Auger-Meitner decay, which corresponds to (x) in Fig.~\ref{fig:Spectra2}(b). Note that there is no (x) peak in the state-resolved approach  in Fig.~\ref{fig:Spectra2}(b). At Ne$^{7+}$, further resonant excitation can happen for both approaches (see Table~\ref{tab:Res} in Appendix~\ref{Tables}). The resulting single-core-hole-excited Ne$^{7+}$ is either $1s^1 2p^1 3p^1$ for the state-resolved approach or $1s^1 3p^1 np^1$ ($n$=$3,4,5$) for the configuration-based approach. The latter can relax via Auger-Meitner decay [peak (xiii) in Table~\ref{tab:AM}], which contributes to the higher yield of Ne$^{8+}$ in the configuration-based approach in Fig.~\ref{fig:Ecomparison}(b). However, for the former case, it is most likely that $1s^1 2p^1 3p^1$ has a $^2P$ state, in which the Auger-Meitner decay is forbidden. Thus, in the state-resolved ionization dynamics, the state of Ne$^{7+} 1s^1 2p^1 3p^1 \ {^2}P$ has to  relax via fluorescence giving rise to peak (v) in Fig.~\ref{fig:Spectra2}(c), which also explains why the yield of Ne$^{8+}$ is suppressed in the state-resolved approach in Fig.~\ref{fig:Ecomparison}(b).
Note that regarding resonances the state-resolved approach is in  better accordance with the findings in Ref.~\cite{Xiang}.
 ~~\\
 
For a short summary, photoelectron, Auger-Meitner electron, and fluorescence spectra provide a plethora of detailed information on x-ray multiphoton ionization dynamics. We demonstrate that resonant excitations and spectral information are described more accurately by the state-resolved implementation due to a general improvement of transition energies and the capture of individual state-resolved features, i.e., transition probabilities.
 In the next section, we explore time-resolved spectra based on the state-resolved Monte Carlo implementation.

\section{Time evolution}\label{Timeevolution}
\begin{figure}[tbp]  
\centering
\includegraphics[trim = 10mm 11mm 5mm 40mm,  clip, width=1\linewidth]{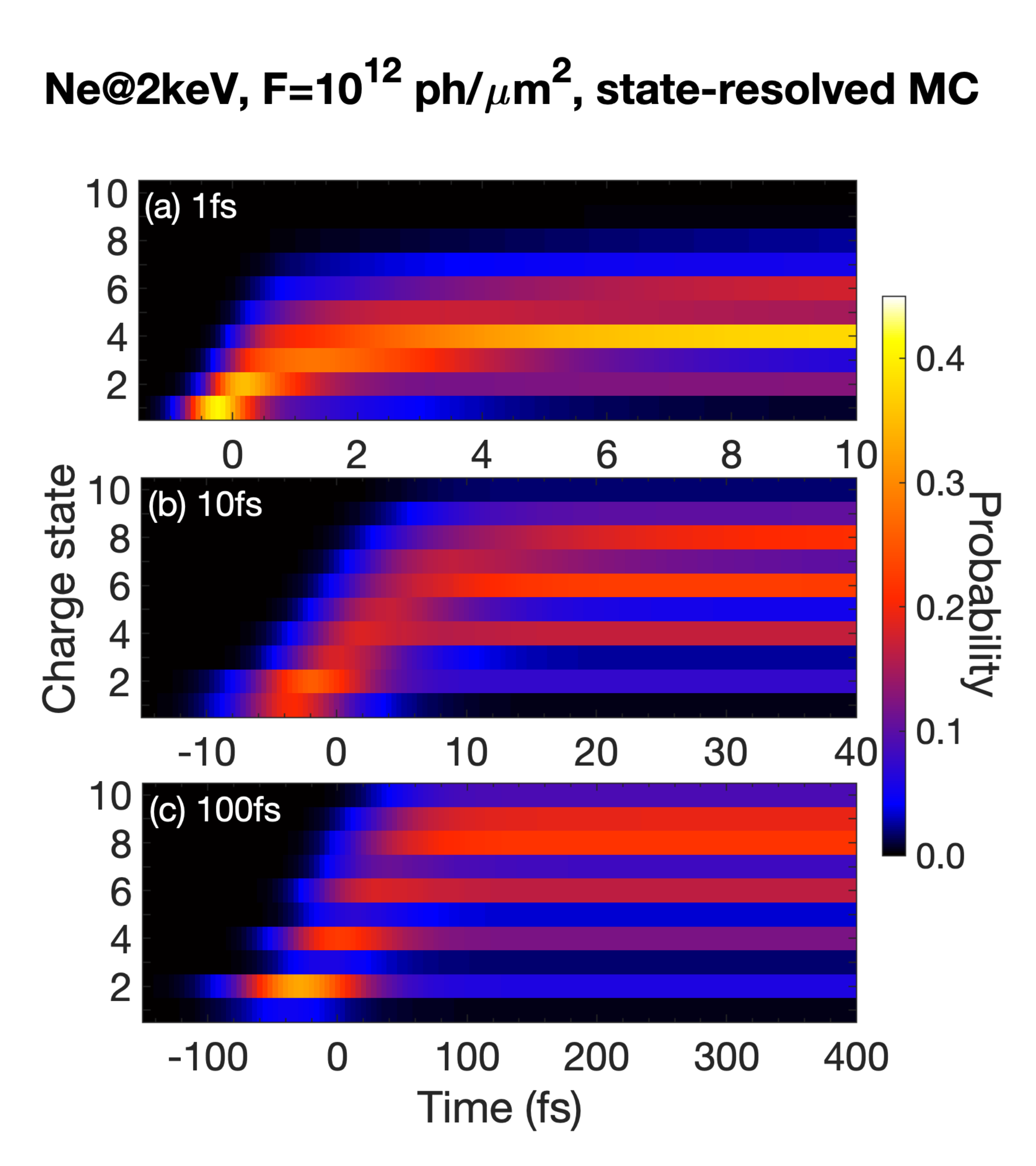}
\caption{Time evolution of Ne CSDs at 2000 eV for pulses duration (FWHM) of (a) 1 fs, (b) 10 fs, and (c) 100 fs. For all cases, the state-resolved Monte Carlo implementation is employed and a fixed fluence of $10^{12}~\text{photons} \slash \mu \text{m}^2$ is used. }
\label{fig:CSDtime}
\end{figure}
\begin{figure}[tbt]  
\centering
\includegraphics[trim = 0mm 0mm 0mm 0mm,  clip, width=1\linewidth]{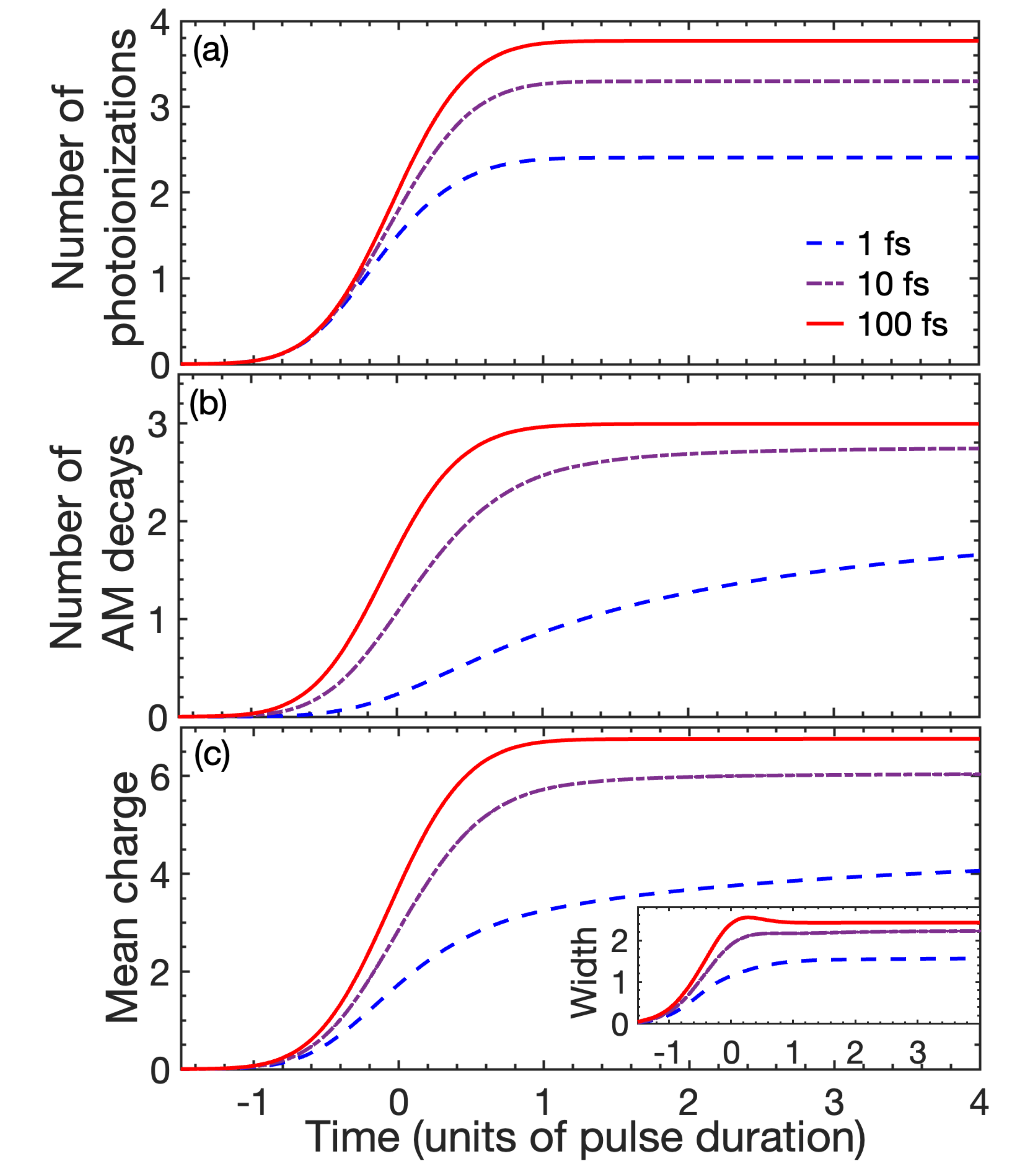}
\caption{Mean number of events for (a) photoionization and (b) Auger-Meitner decay in Ne at 2000~eV as a function of time. (c) Mean charge  and width (inset) of the time-dependent CSDs of Ne shown in Fig.~\ref{fig:CSDtime}. For a better comparability, the time relative to the pulse duration (FWHM) is considered on the $x$ axis. }
\label{fig:Meantime}
\end{figure}
\begin{figure*}[htbp]  
\centering
\includegraphics[trim = 0mm 0mm 0mm 0mm,  clip, width=1\linewidth]{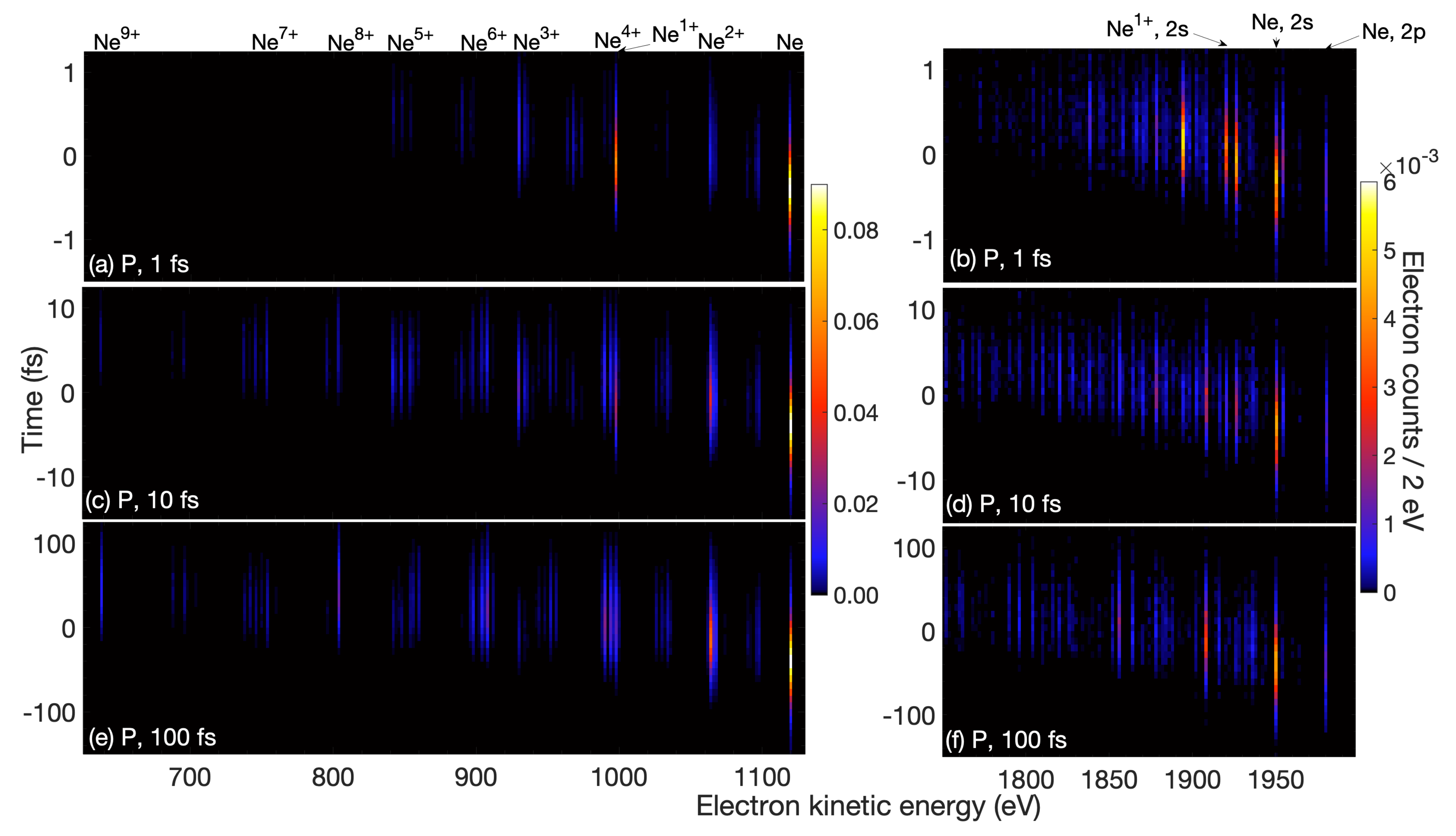}
\caption{Time-resolved photoelectron spectra (P) of Ne at a photon energy of 2000 eV for  pulse durations (FWHM) of (a)-(b) 1~fs, (c)-(d) 10~fs, and (e)-(f) 100~fs.  In (a), (c), and (e) the peaks belong to $1s$ ionization, while those in (b), (d), and (f) belong to $2s$ and $2p$ ionization. The peaks are labelled by the involved initial ion, i.e., \ion{Ne}{q}:  electronic configuration
$1s^2 2l^{8-q}$ for even charges or $1s^1 2l^{9-q}$ for odd charges. 
A fluence of $10^{12}~\text{photons} \slash \mu \text{m}^2$ is used. }
\label{fig:Sptime}
\end{figure*}

\begin{figure}[htbp]  
\centering
\includegraphics[trim = 0mm 0mm 0mm 0mm,  clip, width=1\linewidth]{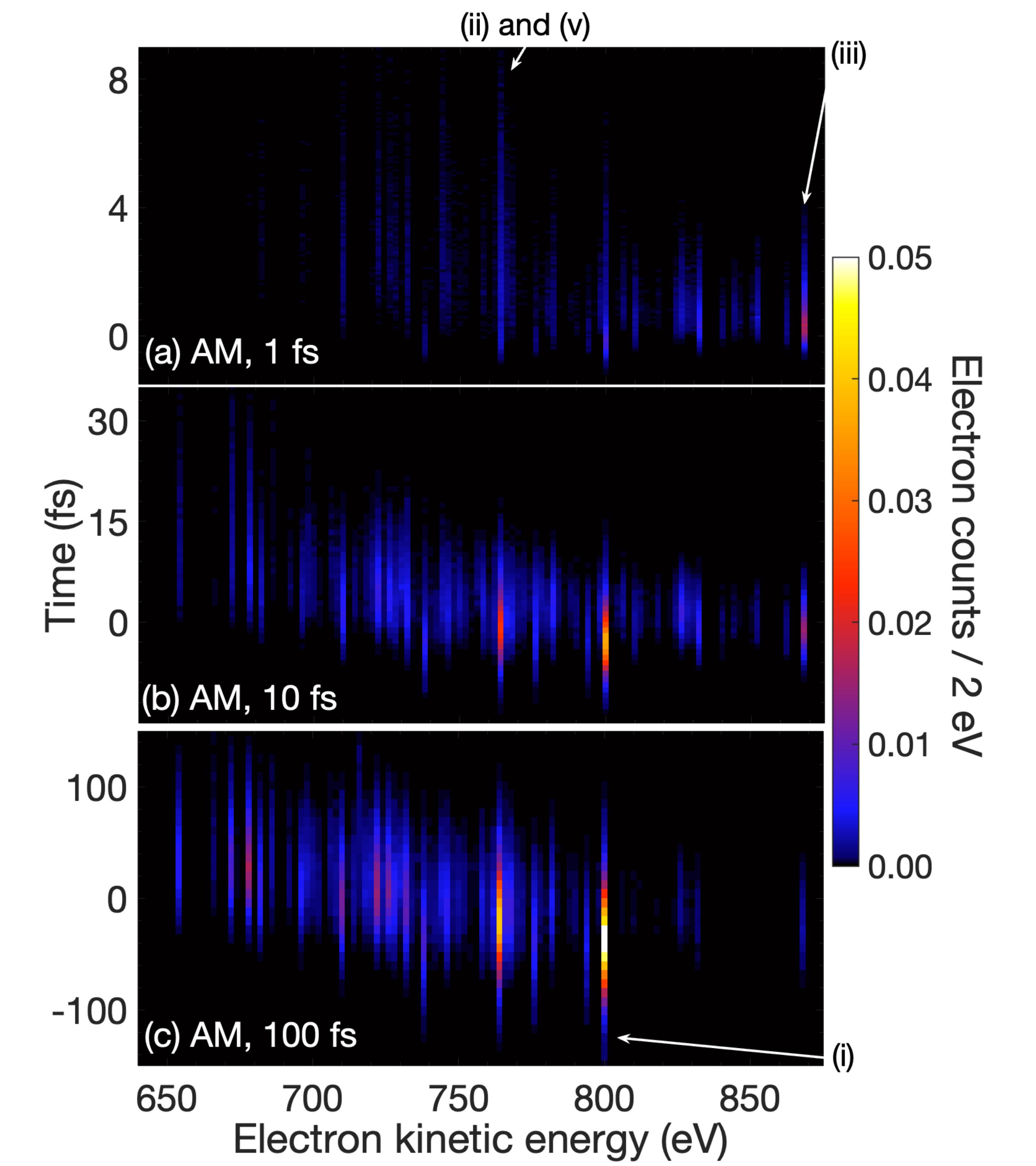}
\caption{Time-resolved Auger-Meitner electron spectra (AM) of Ne at a photon energy of 2000 eV for pulse durations (FWHM) of (a) 1~fs, (b) 10~fs, and (c) 100~fs.   Some peaks are labelled by the transitions listed in Table~\ref{tab:AM} in Appendix~\ref{Tables}. A fluence of $10^{12}~\text{photons} \slash \mu \text{m}^2$ is used. }
\label{fig:Sp2time}
\end{figure}
\begin{figure}[tbp]  
\centering
\includegraphics[trim = 0mm 0mm 0mm 0mm,  clip, width=1\linewidth]{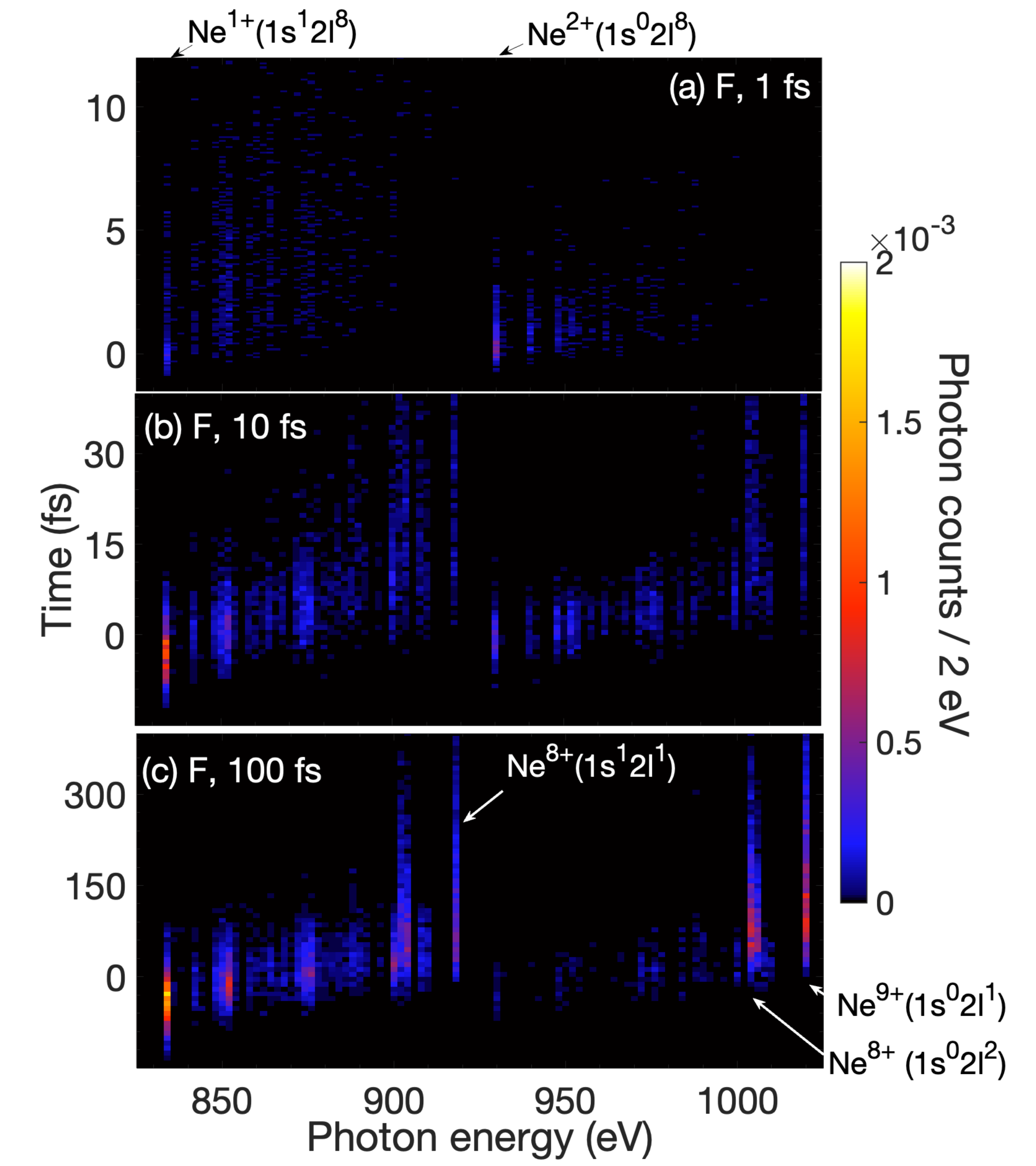}
\caption{Time-resolved $2p\rightarrow 1s$ fluorescence spectra (F) of Ne at  a photon energy of 2000 eV for pulse durations (FWHM) of (a) 1~fs, (b) 10~fs, and (c) 100~fs.  Some peaks are labelled by the involved initial configurations. A fluence of $10^{12}~\text{photons} \slash \mu \text{m}^2$ is used. }
\label{fig:Sp3time}
\end{figure}

We employ the state-resolved Monte Carlo implementation to examine the time evolution of CSDs and spectra for different pulse durations.
We choose pulse durations of 1~fs, 10~fs, and 100~fs FWHM, covering the range of typical pulse durations for current XFEL facilities~\cite{Pellegrini2, Prat, Emma, Tanaka,Kang, Decking}. 
  We consider Ne at a photon energy of 2000~eV and a fluence of $F = 10^{12}~\text{photons} \slash \mu \text{m}^2$, so that in principle all electrons can be ionized via x-ray sequential multiphoton ionization, i.e., a repeated sequence of one-photon ionization and inner-shell relaxation events. In this case, no resonant excitation is involved in the ionization dynamics.

\subsection{Time evolution of charge-state distribution}\label{TimeevolutionCSD}
Figure \ref{fig:CSDtime} presents the Ne CSDs  as a function of time for pulse  durations of (a) 1~fs,  (b) 10~fs, and (c) 100~fs FWHM. The temporal peak is located at $t$=0~fs.  For clarity, the population of neutral Ne is not included. 
Note that the sum of charge populations, including the neutral atom, is unity at every time step. The charge distribution building up looks discrete, which indicates that the populations of odd charges are much smaller than those of even charges. This is a consequence of the fact that   Auger-Meitner decay follows inner-shell photoionization, when the pulse duration is sufficiently longer than the Auger-Meitner lifetime.
One can see that the time-dependent  CSD is shifted to lower charges for a shorter pulse duration. 
For a short pulse duration (1~fs), the CSD peaks around $+4$ and  highly charged ions are barely found as shown in Fig.~\ref{fig:CSDtime}(a). It can be clearly seen that the charged ions are formed sequentially as time goes by, especially near the peak of the pulse. For a long pulse duration (100~fs), however,  the distribution becomes broader with a less pronounced peak at $+8$ as shown in Fig.~\ref{fig:CSDtime}(c). 
These observations are all indicative of frustrated absorption~\cite{Hoener} or intensity-induced x-ray transparency~\cite{Young2}. The degree of ionization is reduced for shorter pulse duration (i.e., higher intensity) because $1s$ photoionization defeats Auger-Meitner decay as the intensity increases. This has two consequences. First,  a double-core-hole state is formed and, thus, the  $1s$ photoionization cross section is reduced (it is zero for $1s^0$).
In Fig.~\ref{fig:Meantime}(a), the mean number of photoionization events is depicted as a function of time for the three pulse durations. It may be seen  that the photoionization number decreases as the pulse duration becomes shorter.
At the same time, the suppression of ionization is also caused by  the reduction of the number of Auger-Meitner decays, depicted in Fig.~\ref{fig:Meantime}(b). These two  mechanisms are responsible for the decreased mean charge [Fig.~\ref{fig:Meantime}(c)] as the pulse is decreased.

Another interesting  observation here is that most  changes in the time-dependent CSD take place within a time interval of $\pm1\times$FWHM.
However, the shorter the pulse, the more extended is the time interval needed to reach the final charge, because the Auger-Meitner lifetime is often tens of fs (see more details in Sec.~\ref{Timeevolutionspectra}).  
In Figs.~\ref{fig:Meantime}(b) and \ref{fig:Meantime}(c), the 1-fs curve is not converged to the temporally asymptotic mean value even at $4\times$FWHM, in contrast to the longer pulse durations.
Therefore, in Fig.~\ref{fig:CSDtime}(a), a longer time interval is considered for the 1-fs result.

\subsection{Time-resolved electron and photon spectra}\label{Timeevolutionspectra}
In order to complete our understanding of the x-ray multiphoton ionization dynamics, we calculate time-resolved  photoelectron (Fig. \ref{fig:Sptime}), Auger-Meitner electron (Fig.~\ref{fig:Sp2time}), and  fluorescence spectra (Fig.~\ref{fig:Sp3time}) for all three pulse durations.  For all figures, the vertical axis is the time, using $0.08\times$FWHM bins, while the horizontal axis is the electron kinetic energy (Figs.~\ref{fig:Sptime} and \ref{fig:Sp2time}) or the photon energy (Fig.~\ref{fig:Sp3time}), using 2-eV bins. 
Note that all the spectra showcase the number of emitted electrons or photons in a time interval relative to the pulse duration since the time binning is adapted for each pulse duration. 

Let us start with the time-resolved photoelectron spectra in Fig.~\ref{fig:Sptime}. 
The $1s$ photoelectron spectra can be grouped according to the peaks belonging to the ionization of \ion{Ne}{q} in a possible configuration $1s^2 2l^{8-q}$ dominantly for even charges or $1s^1 2l^{9-q}$ dominantly for odd charges (with $l=s,p$). Note that the former leads to the formation of single-core-hole states, while double-core-hole states are produced via the latter.  
It is apparent from the spectra that lines corresponding to low odd charges (Ne$^{1+}$ and Ne$^{3+}$) emerge more with shorter pulse duration in Fig.~\ref{fig:Sptime}(a).  The increased number of outer-shell ionizations of lowly charged ions can be observed in Fig.~\ref{fig:Sptime}(b). On the other hand, the lines corresponding to the photoionization of highly charged ions appear more for longer pulse durations as shown in Figs.~\ref{fig:Sptime}(c) and \ref{fig:Sptime}(e), which is consistent with the observation in Fig.~\ref{fig:CSDtime}.

Figure~\ref{fig:Sp2time} shows the time-resolved Auger-Meitner spectra. For long pulse durations, Auger-Meitner decay immediately follows inner-shell ionization and many lines appear in the spectrum as shown in Fig.~\ref{fig:Sp2time}(c).
Here, {\em immediately} is meant relative to the pulse duration, i.e., when the Auger-Meitner lifetime is sufficiently short in comparison to both the pulse duration and the inverse of the resulting inner-shell photoionization rate, so that Auger-Meitner decay can beat further photoionization.  When several Auger-Meitner decays are possible for an initial ion state, lines for more probable processes appear a bit earlier in time.
When the pulse duration is decreased, however, Auger-Meitner decay that takes place on longer time scales than the short pulse duration barely occurs during the pulse. Consequently, the number of processes per time bin is reduced, resulting in weaker lines, covering a longer time range, in Figs.~\ref{fig:Sp2time}(a) and \ref{fig:Sp2time}(b). This reduction of Auger-Meitner decays suppresses refilling of the $1s$ shell and, thus, further inner-shell photoabsorption, which is one of the mechanisms underlying frustrated absorption as discussed in the previous section. 
Note that in our state-resolved calculation the time scales for Auger-Meitner decay for Ne ions are in the range from 818~as (\ion{Ne}{2} $1s^02s^22p^6$) to 46 fs (\ion{Ne}{7} $1s^12s^12p^1$). Thus, most Auger-Meitner decays still take place within 10~fs, as shown in Fig.~\ref{fig:Sp2time}(b).  
Yet another interesting point is that the decay of the double-core-hole state of \ion{Ne}{2}, i.e., peak (iii), is clearly visible for 1~fs and 10~fs, but is almost absent for 100~fs.  This hypersatellite line is located at the highest energy and is well separated from other lines, which provided direct evidence of double-core-hole formation~\cite{Young2, Southworth, Goldsztejn}. Decays of other double-core-hole states, with lower energies than peak (iii),  can also be observed mainly for short pulse durations.

Finally, we turn to the fluorescence spectra for inner-shell relaxation via $2p \rightarrow1s$ transition as shown in Fig.~\ref{fig:Sp3time}. 
We do not show the $2p \rightarrow 2s$ fluorescences spectra that are mainly generated long after the pulse on time scales up to $\sim 10$~ns. For the $2p \rightarrow1s$ fluorescence, we can make very similar observations as for the Auger-Meitner spectra, even though Auger-Meitner decay is much more dominant. However, due to lower fluorescence rates in comparison with Auger-Meitner rates, relaxation of highly charged ions via fluorescence takes place on relatively longer time scales even beyond that shown in Fig.~\ref{fig:Sp3time}. 
Interestingly,  single-core-hole and double-core-hole spectra for Ne ions are well separated and ordered by charge, i.e., the higher the charge state, the higher the photon energy for a fixed number of core electrons. (An analogous effect was observed in XFEL experiments on warm dense aluminum~\cite{Vinko}.) 
The $1s$--$2p$ fluorescence energy is given by $\omega_\text{fluo} = E_I - E_F$, where the initial state $I$ has one or two $1s$ holes and the final state $F$ has one $1s$ hole less than $I$. When $I$ has a double $1s$ vacancy, $E_I$ contains a strong Coulomb repulsion penalty because the two $1s$ holes are spatially close to each other. Thus, the fluorescence lines from double-core-hole states are higher than those from single-core-hole states corresponding to the same charge state. As the charge state increases, both $E_I$ and $E_F$ increase, but the increase is less for $E_F$ than $E_I$ because in state $F$ there is more screening by $1s$ electrons than in state $I$. Consequently, the fluorescence energy gets larger for higher charge states, which are generated at later times. All these features are demonstrated in the time-dependent fluorescence spectra: In each panel of Fig.~\ref{fig:Sp3time}, there are two groups of transition lines---associated with single- and double-core-hole states, respectively---that move toward higher energies with increasing time.

\section{Conclusion}\label{conclusion}
In this paper, we have presented an implementation of  state-resolved Monte Carlo calculations for describing x-ray multiphoton ionization dynamics. Our implementation in the \textsc{xatom} toolkit employs an improved electronic-structure calculation that is based on first-order many-body perturbation theory.
We have compared the new state-resolved and the original configuration-based Monte Carlo calculations for neon at three different photon energies, including a resonant case.
The differences in the CSD  are negligible when resonances do not matter. 
Therefore, in these cases the original configuration-based version of \textsc{xatom} already produces quite good results as demonstrated in former studies~\cite{Doumy,Rudek2, Rudek, Rudek3, Motomura, Fukuzawa}.
 However, resonant excitations and spectral informations  profit from the improved first-order-corrected transition energies in the new implementation. Our results have demonstrated that CSD for the resonance case and calculated electron and photon spectra are improved by using state-resolved ionization dynamics calculations. 
Employing the state-resolved Monte Carlo implementation, we have investigated CSDs and spectra of neon atoms at a photon energy of 2000~eV  for three different XFEL pulse durations. 
In addition to asymptotic quantities, we have computed time-resolved CSDs and spectra, which highlight the mechanisms through which different pulse durations affect the asymptotic observables.  In our example, frustrated absorption clearly manifests itself in the time-resolved spectra as the pulse duration gets shorter. Particularly, it is the dynamical interplay between the suppression of Auger-Meitner decay and the suppression of inner-shell photoabsorption that lies at the heart of {\em frustrated absorption}---not the suppression of photoabsorption alone.
 
We conclude with an outlook.
An important next step could be to employ our state-resolved Monte Carlo implementation to explore the orbital alignment of the ions~\cite{TIofOA,Gryzlova} produced during the x-ray multiphoton ionization  dynamics.
Another promising perspective for further developments is to compare our results on time-dependent quantities with experimental measurements. Experimental methods that may allow such measurements could be attosecond transient absorption spectroscopy~\cite{Goulielmakis, Wang} and attosecond streaking measurements~\cite{Itatani,Hentschel, Krausz}. These methods have already been employed to investigate ionization dynamics under conditions in which only a few processes were involved~\cite{Drescher, Schultze}. Probing dynamics during x-ray multiphoton ionization may be experimentally challenging due to a wide variety of involved charge states and emitted photo- and Auger-Meitner electrons that are not always well separated, neither in time nor in energy.

\section{Acknowledgements}\label{Acknowledgements}
We acknowledge the support by DASHH (Data Science in Hamburg - HELMHOLTZ Graduate School for the Structure of Matter) with the Grant-No.\ HIDSS-0002.~~\\

\appendix
\section{Physical processes and transition energies}\label{Tables}
\begin{table*}[htbp]
\centering
\caption{Ionization potentials of selected processes calculated with 
the state-resolved approach, $E_\text{IP}^{(1)}$, and the configuration-based approach, $E_\text{IP}^{(0)}$. Photoemission lines in Figs.~\ref{fig:Spectra1}(a), \ref{fig:Spectra3}(a), and \ref{fig:Spectra2}(a) are assigned by $E = \omega_\text{in} - E_\text{IP}$ and their labels are listed below.  }
\label{tab:P}
\begin{ruledtabular}
\begin{tabular}{ccrr}
label&process& $E_\text{IP}^{(1)} $(eV) & $E_\text{IP}^{(0)} $ (eV)\\
 \midrule
 (i)&~~Ne, $1s^22s^22p^6 \rightarrow 1s^12s^22p^6  $ & 880 ($^1S \rightarrow {^2}S$)
&857\\[4pt]
  (ii)&\ion{Ne}{2}, $1s^22s^22p^4  \rightarrow 1s^12s^22p^4  $ & 935 ($^1D \rightarrow {^2}D$)&912\\[4pt]
  (iii)&\ion{Ne}{1}, $1s^12s^22p^6  \rightarrow 1s^02s^22p^6  $ & 1001 ($^2S \rightarrow {^1}S$)&952\\[4pt]
   (iv)&\ion{Ne}{4}, $1s^22s^12p^3  \rightarrow 1s^12s^12p^3  $ & 1006 ($^1D \rightarrow {^2}D$)&983\\[4pt]
    (v)&\ion{Ne}{4}, $1s^22s^22p^2  \rightarrow 1s^12s^22p^2  $ & 1010 ($^1D \rightarrow {^2}D$)&986\\[4pt]
     (vi)&\ion{Ne}{3}, $1s^12s^22p^4  \rightarrow 1s^02s^22p^4  $ & 1070 ($^2D \rightarrow {^1}D$)&1020\\[4pt]
     (vii)&~~ Ne, $1s^22s^22p^6  \rightarrow 1s^22s^12p^6  $ & 49 ($^1S \rightarrow {^2}S$)&43\\[4pt]
   (viii)&~~Ne, $1s^22s^22p^6  \rightarrow 1s^22s^22p^5  $ & 20 ($^1S \rightarrow {^2}P$)&20\\[4pt]
  (ix)&\ion{Ne}{1}, $1s^22s^12p^6  \rightarrow 1s^22s^02p^6  $ & 72 ($^2S \rightarrow {^1}S$)&64\\[4pt]
  (x)&\ion{Ne}{1}, $1s^22s^22p^5  \rightarrow 1s^22s^12p^5 $ & 66 ($^2P \rightarrow {^3}P$)&64\\  
   &~~~~~~~~~~~~~~~~~~~~~~~~~~~ $  $ & 77 ($^2P \rightarrow {^1}P$)&--\\       
\end{tabular}
\end{ruledtabular}
\end{table*}
\begin{table*}[htbp]
\centering
\caption{Peak assignment in the Auger-Meitner electron spectra [Figs.~\ref{fig:Spectra3}(b) and \ref{fig:Spectra2}(b)]. 
Transition energies for the state-resolved approach, $E_\text{tr}^{(1)}$, and the configuration-based approach, $E_\text{tr}^{(0)}$, are listed for the underlying process. }
\label{tab:AM}
\begin{ruledtabular}
\begin{tabular}{ccrr}
label&process& $E_\text{tr}^{(1)}$ (eV) & $E_\text{tr}^{(0)} $ (eV)\\
 \midrule
 (i)&\ion{Ne}{1}, $1s^12s^22p^6  \rightarrow 1s^22s^22p^4  $ & 801($^2S \rightarrow {^1}D$)&860\\
 &~~~~~~~~~~~~~~~~~~~~~~~~~~~  $$ & 795 ($^2S \rightarrow {^1}S$)&--\\[4pt]
  (ii)&\ion{Ne}{1}, $1s^12s^22p^6  \rightarrow 1s^22s^12p^5  $ & 764 ($^2S \rightarrow {^1}P$)&838\\
    &~~~~~~~~~~~~~~~~~~~~~  $ \rightarrow 1s^22s^12p^5  $ & 777 ($^2S \rightarrow {^3}P$)&--\\[4pt]
   (iii)&\ion{Ne}{2}, $1s^02s^22p^6 \rightarrow 1s^12s^22p^4  $ & 869 ($^1S \rightarrow {^2}D$)&875\\
     &~~~~~~~~~~~~~~~~~~~~~~~~~~~ $$ & 863 ($^1S \rightarrow {^2}S$)&--\\[4pt]
        (iv)&\ion{Ne}{2}, $1s^02s^22p^6 \rightarrow 1s^12s^12p^5  $ & 833 ($^1S \rightarrow {^2}P$)&856\\[4pt]
    (v)&\ion{Ne}{3}, $1s^12s^22p^4  \rightarrow 1s^22s^22p^2  $ & 764 ($^2D \rightarrow {^1}D$)&820\\
         &~~~~~~~~~~~~~~~~~~~~~~~~~~~ $$ & 758 ($^2D \rightarrow {^1}S$)& --\\[4pt]
     (vi)&\ion{Ne}{3}, $1s^12s^22p^4  \rightarrow 1s^22s^12p^3  $ & 733 ($^2D \rightarrow {^1}D$)&800\\
              &~~~~~~~~~~~~~~~~~~~~~~~~~~~ $ $ & 729 ($^2D \rightarrow {^1}P$)&--\\[4pt]
   (vii)&\ion{Ne}{4}, $1s^02s^22p^4  \rightarrow 1s^12s^22p^2  $ & 826 ($^1D \rightarrow {^2}D$)&830\\[4pt]
     (viii)&\ion{Ne}{7}, $1s^12s^12p^1  \rightarrow 1s^22s^02p^0 $ & 673 ($^2P \rightarrow {^1}S$)&717\\[4pt]
   (ix)&\ion{Ne}{7}, $1s^12s^22p^0  \rightarrow 1s^22s^02p^0  $ & 654 ($^2S \rightarrow {^1}S$)&706\\ [4pt]
   (x)&\ion{Ne}{5}, $1s^02s^22p^23p^1  \rightarrow 1s^12s^12p^13p^1  $ &-- &788\\[4pt]     
   (xi)&\ion{Ne}{6}, $1s^12p^2np^1   \rightarrow 1s^22p^0np^1   $ &690 ($n=4;\ ^1F \rightarrow {^2}P$) &741 ($n=5$)\\ 
  &~~~~~~~ $ $ & 690 ($n=4;\ ^1D \rightarrow {^2}P$)&739 ($n=6$)\\ 
    &&-- &737 ($n=7$)\\ [4pt]
       (xii)&\ion{Ne}{6}, $1s^24p^2    \rightarrow 1s^22p^1   $ &119 ($^3P \rightarrow {^2}P$)  &125 \\ 
  &~~~~ $ $ & 14 ($^3D \rightarrow {^2}P$) &13\\ [4pt]  
   (xiii)&\ion{Ne}{7}, $1s^13p^1np^1    \rightarrow 1s^12p^1   ~~~$ &43 ($n=3;\ ^2P \rightarrow {^1}P$) &44 ($n=3$)\\ 
  &~~~~~$1s^13p^1np^1   \rightarrow 1s^12p^1  $ & 51 ($n=3;\ ^2D \rightarrow {^3}P$)&119 ($n=5$)\\ 
        &&-- &131 ($n=6$)\\ 
            &&-- &138 ($n=7$)\\ 
\end{tabular}
\end{ruledtabular}
\end{table*}
\begin{table*}[htbp]
\centering
\caption{Peak assignment in the fluorescence spectra [Figs.~\ref{fig:Spectra3}(c) and \ref{fig:Spectra2}(c)]. 
Transition energies for the state-resolved approach, $E_\text{tr}^{(1)}$, and the configuration-based approach, $E_\text{tr}^{(0)}$, are listed for the underlying process. }
\label{tab:F}
\begin{ruledtabular}
\begin{tabular}{ccrr}
label&process& $E_\text{tr}^{(1)}$ (eV) & $E_\text{tr}^{(0)} $ (eV)\\
 \midrule
 (i)&\ion{Ne}{1}, $1s^12s^22p^6  \rightarrow 1s^22s^22p^5  $ & 834 ($^2S \rightarrow {^2}P$)&906\\[4pt]
  (ii)&\ion{Ne}{8}, $1s^12s^02p^1  \rightarrow 1s^22s^02p^0 $ & 919 ($^1P \rightarrow {^1}S$)&972\\[4pt]
  (iii)&\ion{Ne}{8}, $1s^12s^02p^1  \rightarrow 1s^12s^12p^0 $ & 10 ($^3P \rightarrow {^3}S$)&11\\
       & & 7 ($^1P \rightarrow {^1}S$)&--\\[4pt]
   (iv)&\ion{Ne}{9}, $1s^02s^02p^1  \rightarrow 1s^12s^02p^0 $ & 1020 ($^2P \rightarrow {^2}S$)&1020\\ [4pt]     
   (v)&\ion{Ne}{7}, $1s^12p^13p^1  \rightarrow 1s^22p^03p^1  $ &916 ($^2P \rightarrow {^2}P$)&--\\  [4pt]    
   (vi)&\ion{Ne}{7}, $1s^2np^1  \rightarrow 1s^22s^1  $ &140 ($n=3;\ ^2P \rightarrow {^2}S$)&143 ($n=3$)\\  
    & &183 ($n=4;\ ^2P \rightarrow {^2}S$)&188 ($n=4$)\\
     & &203 ($n=5;\ ^2P \rightarrow {^2}S$)&210 ($n=5$)\\
     & &214 ($n=6;\ ^2P \rightarrow {^2}S$)&222 ($n=6$)\\
      & &220 ($n=7;\ ^2P \rightarrow {^2}S$)&230 ($n=7$)\\
\end{tabular}
\end{ruledtabular}
\end{table*}
\begin{table*}[htbp]
\centering
\caption{Dominant resonant excitations at a photon energy of 1050 eV ($1\%$ bandwidth). Transition energies and cross sections for the state-resolved approach, $E_\text{tr}^{(1)}$ and $\sigma^{(1)}$, and the configuration-based approach, $E_\text{tr}^{(0)}$ and $\sigma^{(0)}$, are listed for the underlying process.  For brevity,  only a range of transition energies in the state-resolved approach are given instead of individual state-to-state transition energies. For the same reason, only subshell cross sections, i.e., averages over initial states and sums over all final states, are shown. For  \ion{Ne}{3} ($1s^12l^6$), \ion{Ne}{5} ($1s^12l^4$), \ion{Ne}{5} ($1s^22l^24p^1$), \ion{Ne}{6} ($1s^22l^2$), and \ion{Ne}{7} ($1s^22l^1$) with $l=s,p$, similar resonant excitations are also possible for other electronic configurations than given here, but are not listed for brevity.
}
\label{tab:Res}
\begin{ruledtabular}
\begin{tabular}{crrrrr}
process& $n$& $E_\text{tr}^{(1)}$ (eV) & $\sigma^{(1)}$ (Mb) & $E_\text{tr}^{(0)}$ (eV) & $\sigma^{(0)}$ (Mb)\\
 \midrule
 \ion{Ne}{3}, $1s^12s^22p^4 \rightarrow 1s^02s^22p^4np^1$ & 4&1043--1059& $6.98 \times10^{-2}$ &-- &--\\
& 5& 1050--1065& $7.74 \times 10^{-3}$ & --&--\\[4pt]    
 \ion{Ne}{5}, $1s^12s^22p^2 \rightarrow 1s^02s^22p^2np^1$ & 3&--& --& 1043&$1.28 \times 10^{-1}$ \\[4pt]  
  \ion{Ne}{5}, $1s^22s^24p^1 \rightarrow 1s^12s^24p^1np^1$ & 4&1052--1054& $2.73 \times 10^{-1}$& 1031&$2.62 \times 10^{-5}$ \\[4pt]  
  \ion{Ne}{6}, $1s^22s^02p^2 \rightarrow 1s^12s^02p^2np^1$ & 4&1040--1053& $2.85 \times10^{-1}$ &-- &--\\
& 5& 1056--1070& $7.60 \times 10^{-3}$ & 1041 &$2.97 \times10^{-2}$\\
& 6& 1065--1069& $6.19 \times 10^{-3}$ & 1050 &$1.23 \times 10^{-1}$\\
& 7& --& -- & 1055 &$4.32 \times 10^{-2}$\\[4pt] 
 \ion{Ne}{7}, $1s^22s^02p^1 \rightarrow 1s^12s^02p^1np^1$ & 3&1036--1046& $1.73 \times10^{-1}$ &-- &--\\
& 4& --& -- & 1063&$6.26 \times 10^{-3}$\\[4pt]  
 \ion{Ne}{7}, $1s^2np^1 \rightarrow 1s^13p^1np^1$ & 3&1064--1067& $9.51 \times10^{-3}$ & 1047 &$8.67 \times 10^{-1}$\\
& 4& 1067--1070& $1.34 \times 10^{-4}$ & 1056 &$4.84 \times 10^{-1}$\\
& 5& 1070& $2.44 \times 10^{-6}$ & 1061 &$7.07 \times 10^{-2}$\\
& 6& 1070& $3.14 \times 10^{-7}$ & 1063 &$1.77 \times 10^{-2}$\\
& 7& 1070& $7.67 \times 10^{-8}$ & 1064 &$6.85 \times 10^{-3}$\\
\end{tabular}
\end{ruledtabular}
\end{table*}

For dominant physical processes, calculated ionization potentials and transition energies are listed in Tables~\ref{tab:P}--\ref{tab:Res}. 


\end{document}